\documentstyle[aps,epsfig,prl]{revtex}
\begin{document}
\title{The Quantum Theory of Conductivity of Spatially - Heterogeneous Systems}
\author{I. G. Lang, L. I. Korovin}
\address{A. F. Ioffe Physico-technical Institute, Russian
Academy of Sciences, 194021  St. Petersburg, Russia}
\author{J. A. de la Cruz-Alcaz\dag, S. T. Pavlov\dag\ddag}
\address{\dag Facultad de Fisica de la UAZ, Apartado Postal C-580,
98060 Zacatecas, Zac., Mexico \\
\ddag P. N. Lebedev Physical Institute , Russian Academy of Sciences, 119991 Moscow,
Russia}

 \twocolumn[\hsize\textwidth\columnwidth\hsize\csname @twocolumnfalse\endcsname
\maketitle\widetext
\begin{abstract}
\begin{center}
\parbox {6in} {
The quantum theory of conductivity of semiconductor objects, to which the quantum
wells, wires and dots concern, is constructed. Average values of
  current and charge densities, induced
by a weak electromagnetic field, are calculated. It is shown,
that in both cases average current and charge densities contain
two contributions, first of which is expressed through electric
field, and second - through a spatial derivative of electric
field.  Appropriate expressions for the conductivity tensor,
dependent on coordinates and applicable to any
spatially-heterogeneous systems, are deduced. The results may be
used in the theory of secondary light radiation from
low-dimensional objects in cases of monochromatic light and light
pulses.}
\end{center}
\end{abstract}

] \narrowtext

\section{Introduction}
In connection with the increased interest to experiment and theory concerning  light
reflection and absorption by low-dimensional semiconductor objects  - quantum wells,
wires  and dots - at light pulse excitation (see, for example, \cite{1,2}) there
appears a question on what kind of electromagnetic wave-electron interaction is more
convenient to use, - containing vector potential $ {\bf A} ({\bf r}, t) $ or electric
field $ {\bf E} ({\bf r}, t). $ The work \cite{3} is devoted to the same question
with reference to calculation of a differential section of inelastic light scattering
by infinite crystals. In \cite{3} it is shown, that exact expressions for scattering
sections (with use sets of exact electron wave functions in a crystal), obtained with
use of two various kinds of electromagnetic wave-electron interaction, coincide. But
the set of exact electron wave functions in a crystal (taking into account precisely,
for example, electron-phonon interaction) is unknown, therefore at calculation of
sections the approached methods are used. Namely, it is taken into account
electron-phonon interaction in the lowest order,  not all intermediate states of an
electronic system are considered. If to use these approximations, various kinds of
light-electron interaction (containing $ {\bf A} $ or $ {\bf E} $) will result in
distinguished results. The authors of\cite{3} approve, that in a case of non-resonant
scattering the use of interaction containing $ {\bf E} $, gives the best results.

We assume to construct the general theory of secondary light radiation  by
low-dimensional semiconductor objects, to which number quantum wells, quantum wires
and quantum dots concern. First of all, the theory should describe light reflection
and absorption
 by such objects, and also various kinds of  light  scattering (Raman
scattering, Releigh scattering). The theory should be applicable in a case of
monochromatic irradiation, as well as in a case of a light pulse irradiation. We are
limited to linear approximation on intensity of exciting light. From told above
follows, that others tasks are decided, than in \cite{3}, where light scattering in
bulk crystals is considered. Therefore the problem of an interaction choice (through $
{\bf A} $ or $ {\bf E} $) is necessary to be solved anew.

In the present work we shall calculate average values  of  current and charge
densities induced by an electromagnetic field  in the case of spatially-heterogeneous
matter. To this case concern semiconductor objects of the lowered dimension. Having
calculated linear on electric and magnetic fields the contributions to average values
of  current and charge densities, we can then to determine these fields inside and
outside of semiconductor objects, solving the Maxwell equations. Expressions for
fields appropriate  reflected and past through object light thus can be obtained. Such
procedure, taking into account all orders light-electron interaction,
 is done in \cite{4}, where intensity of
reflected and absorbed light at monochromatic irradiation quantum well of final
thickness are calculated . In the present work we deduce expression for average
induced  current density used in\cite{4}.

The operator of interaction of the charged particles with electric and magnetic
fields is expressed through vector $ {\bf A} ({\bf r}, t) $ and scalar $ \varphi ({\bf
r}, t) $ potentials. Therefore average values of the induced of current and charge
densities also are expressed through these potentials. However to use these
expressions is inconvenient because of the contribution
$$- {e\over mc} \langle0 |\rho ({\bf r}) |0\rangle A_\alpha ({\bf r}, t). $$
 In average of a current density , where $e, m $ is the charge and mass of a particle
respectively, $ \rho ({\bf r}) $ is the operator  of a charge density (see section IV
lower). Therefore we shall express average values  of  current and charge densities
through electric and magnetic fields. It is possible since all observable values are
expressed through  electric $ {\bf E} ({\bf r}, t) $ and magnetic $ {\bf H} ({\bf r},
t) $ fields. Our task will consist in transition from expressions for physical values,
containing potentials, to expressions containing fields $ {\bf E} ({\bf r}, t) $ and
$ {\bf H} ({\bf r}, t). $

In the present work the case of temperature $T=0 $ is considered. Further we assume
to calculate the average induced current and charge densities in a case spatial -
heterogeneous systems and final temperatures. We shall use some results of
\cite{8,9,10}.

The article is organized as follows. In sections II-IV the
statement of a task is stated, operators of current and charge
densities and their average values on ground state of system are
entered. In sections V -IX the task about expression of average
values through an electric field and its spatial derivatives is
solved. Section X is devoted to exception of  average values of
diagonal matrix elements of operators $ {\bf r} _i $. In section
XI the general expression for conductivity tensor of
heterogeneous system is given. In sections XII and XIII the case
of zero electric and constant magnetic fields is considered.

\section{Statement of a task.}

Let us consider s system of $N $ particles with the mass $m $ and charge $e $ in a
weak electromagnetic field characterized by intensities $ {\bf E} ({\bf r}, t) $ and
$ {\bf H} ({\bf r}, t) $. Let us introduce the vector $ {\bf A} ({\bf r}, t) $ and
scalar $ \varphi ({\bf r}, t) $ potentials
\begin{eqnarray}
\label {1} {\bf E} ({\bf r}, t)&=&-{1\over c} {\partial {\bf A} \over\partial t}
-\nabla
\varphi, \nonumber\\
 {\bf H} ({\bf r}, t)&=&\nabla\times{\bf A}.
\end{eqnarray}
We shall consider fields as classical. The gage of $ {\bf A} ({\bf r}, t) $ and $
\varphi ({\bf r}, t) $ is arbitral. For completeness of a task we shall consider,
that the system of particles can be placed in a constant magnetic field $ {\bf H} _c
$, which may be strong (SMF). This field is described by the vector potential $ {\cal
A} ({\bf r}) $, so
$$ {\bf H} _c=rot \, {\cal A} ({\bf r}). $$
The total Hamiltonian $ {\bf {\cal H}} _ {total} $ is written  as
\begin{eqnarray}
\label {2} {\bf {\cal H}} _ {total} = {1\over 2m} \sum_i\left ({\bf {\cal {P}}} _i-
{e\over c} {\bf
{ \cal {A}}} ({\bf r} _i) - {e\over c} {\bf A} ({\bf r} _i, t) \right) ^2\nonumber\\
+V ({\bf r} _1\ldots {\bf r} _N) +e\sum_i\varphi ({\bf r} _i, t),
\end{eqnarray}
where $ {\bf {\cal {P}}} _i =-i\hbar (\partial/\partial {\bf r} _i) $ is the
generalized momentum operator (see, for example, \cite{5,6}), $V ({\bf r} _1\ldots
{\bf r} _N) $ is the potential energy including interaction between particles and the
external potential. In Eq. (2) It is necessary to take into account non-commutativity
of $ {\bf {\cal P}} _i $ and $ {\cal A} ({\bf r} _i), {\bf A} ({\bf r} _i, t) $. Let
us allocate in Eq. (2) energy $U $ interaction of particles with the electromagnetic
field, having included interaction with a strong magnetic field in the main
Hamiltonian
\begin{equation}
\label {3} {\cal {H}} _ {total} = {\cal {H}} +U,
\end{equation}
\begin{equation}
\label {4} {\cal {H}} =
 {1\over 2m} \sum_i {\bf p} _i^2+V ({\bf r} _1\ldots {\bf r} _N), ~~
{ \bf p} _i = {\bf {\cal {P}}} _i- {e\over c} {\bf {\cal {A}}} ({\bf r} _i),
\end{equation}
\begin{eqnarray}
\label {5} U=U_1+U_2, \nonumber\\
U_1 =- {1\over c} \int d^3r {\bf j} ({\bf r}) {\bf A} ({\bf r},
t) + \int d^3r  \rho ({\bf
r}) \varphi ({\bf r}, t), \nonumber\\
U_2 = {e\over 2mc} \int d^3r \rho ({\bf r}) {\bf A} ^2 ({\bf r}, t),
\end{eqnarray}
where the operators of  current and charge densities are introduced
\begin{eqnarray}
\label {6} {\bf j} ({\bf r}, t) = \sum_i {\bf j} _i ({\bf r}, t), \nonumber\\
 {\bf j} _i ({\bf r}, t) = {e\over 2} \{\delta ({\bf r} - {\bf r} _i) {\bf v} _i
 + {\bf v} _i\delta ({\bf r} - {\bf r} _i)\}, ~~ {\bf v} _i = {\bf p} _i/m,\nonumber\\
 \rho ({\bf r}) = \sum_i\rho _i ({\bf r}), ~~~\rho_i ({\bf r}) =e\delta ({\bf r} - {\bf r} _i).
\end{eqnarray}
Our task is to calculate in linear approximation on external fields $ {\bf E} ({\bf
r}, t) $ and $ {\bf H} ({\bf r}, t) $ the induced current and charge densities
averaged on the system  ground state.

\section{The definition of operators}
The charge density operator $ \rho ({\bf r}) $ in the Schr\"odinger  representation
does not contain  additives proportional to fields, but the current density operator
in fields looks like
$$ {\bf j} ({\bf r}, t) + \Delta {\bf j} ({\bf r}), $$
where
\begin{equation}
\label {7} \Delta {\bf j} ({\bf r}) = {e\over 2} \sum_i \{\Delta {\bf v} _i\delta
({\bf r} - {\bf r} _i) + \delta ({\bf r} - {\bf r} _i) \Delta {\bf v} _i \},
\end{equation}
\begin{equation}
\label {8} \Delta {\bf v} _i = (i/\hbar) [U, {\bf r} _i] = - {e\over mc} {\bf A}
({\bf r} _i, t),
\end{equation}
$ [F, Q] $ is the commutator of operators $F $ and $Q $, and hence
\begin{equation}
\label {9} \Delta {\bf j} ({\bf r}) = - {e\over mc} \rho ({\bf r}) {\bf {A}} ({\bf
r}, t).
\end{equation}
In the interaction representation we have
\begin{eqnarray}
\label {10} \rho ({\bf r}, t)&=&e^{i{\cal{H}}t/\hbar}\rho({\bf
r}) e ^ {-i {\cal {H}} t/\hbar}, \nonumber\\
 {\bf j} ({\bf r}, t)&=&e^{i{\cal{H}}t/\hbar}{\bf
j} ({\bf r}) e ^ {-i {\cal {H}} t/\hbar}, \nonumber\\
\Delta {\bf j} ({\bf r}, t)&=&-{e\over mc} \rho ({\bf r}, t) {\bf {A}} ({\bf r}, t).
\end{eqnarray}

 Now we shall determine the current and charge density operators
  in the Heisenberg representation. In
 \cite{7} (page 82) it is shown, that connection between the operator $F (t) $
  in the interaction representation and
 the operator $F_G (t) $ in the Heisenberg representation is expressed as
\begin{equation}
\label {11} F_G (t) =S ^ {-1}(t) F (t) S (t),
\end{equation}
where $S$ is the S-matrix. It is determined as
\begin{eqnarray}
\label {12} S(t)&=&S(t,-\infty)=1-
{ i\over\hbar}\int^t_{-\infty}dt_1U(t_1)\nonumber\\
 &+&\left(-{i\over\hbar}\right)^2
 \int^t_{-\infty}dt_2U(t_2)\int^{t_2}_{-\infty}dt_1U(t_1)+\ldots,
\end{eqnarray}
\begin{equation}
 \label{11}U(t)=e^{i{\cal{H}}t/\hbar}Ue^{-i{\cal{H}}t/\hbar}.
\end{equation}
Using Eq. (12), we find, that linear additives on potentials $ {\bf A} ({\bf r}, t) $
and $ \varphi ({\bf r}, t) $  to the current and charge density  operators in the
Heisenberg representation are equal
$$ j _ {1\alpha} ({\bf r}, t) = \Delta j _ {\alpha} ({\bf r}, t) - {i\over\hbar} \int^t _ {-\infty}
dt ^\prime [j _ {\alpha} ({\bf r}, t), U_1 (t ^\prime)], ~~ (13a) $$
$$\rho _ {1} ({\bf r}, t) = - {i\over\hbar} \int^t _ {-\infty}
dt ^\prime [\rho ({\bf r}, t), U_1 (t ^\prime)]. ~~~~~ (13b) $$ The index 1 means the
first order on potentials $ {\bf A} ({\bf r}, t) $ and $ \varphi ({\bf r}, t) $.
Substituting Eqs. (9) and (5) in last expressions, we obtain
\begin{eqnarray}
\label {14} j _ {1\alpha} ({\bf r}, t) = - {e\over mc} \rho ({\bf r}, t) A_\alpha
({\bf r},
t) \nonumber\\
+ {i\over \hbar c} \int d^3r^\prime\int_{-\infty}^tdt^\prime[j_\alpha ({\bf r}, t),
j_\beta ({\bf
r} ^ \prime, t ^\prime)] A_\beta ({\bf r} ^ \prime, t ^\prime) \nonumber\\
- {i\over \hbar} \int d^3r^\prime\int_{-\infty}^tdt^\prime[j_\alpha ({\bf r}, t), \rho
( {\bf r} ^ \prime, t ^\prime)] \varphi ({\bf r} ^ \prime, t ^\prime),
\end{eqnarray}
\begin{eqnarray}
\label {15} \rho_1 ({\bf r}, t)&=&{i\over \hbar c} \int
 d^3r^\prime\int_{-\infty}^tdt^\prime[\rho ({\bf r}, t), j_\beta ({\bf
r} ^ \prime, t ^\prime)] A_\beta ({\bf r} ^ \prime, t ^\prime) \nonumber\\
 &-&{i\over \hbar} \int d^3r^\prime\int_{-\infty}^tdt^\prime[\rho ({\bf r}, t), \rho
( {\bf r} ^ \prime, t ^\prime)] \varphi ({\bf r} ^ \prime, t ^\prime).
\end{eqnarray}

\section{Averaging on the system ground state}
Let us consider a zero temperature case  and average Eqs. (14), (15) on the ground
state of our system. In all further calculations we shall assume, that {\it on the
indefinitely removed distances there are no charges and currents, and also that on
times}~$ t\to -\infty $  $ {\bf E} ({\bf r}, t) $ {\it and} $ {\bf H} ({\bf r}, t) $
{\it are equal 0}, what corresponds to an adiabatic switching-on of these fields. In
\cite{7} (page 84) it is shown, that at averaging it is necessary to use wave
functions $|0\rangle$ of the ground state without taking into account the interaction
$U$. For averaged values of current and charge densities we shall enter designations
$ \langle 0 | {\bf j} _1 ({\bf r}, t) |0\rangle $ and $ \langle 0 |\rho_1 ({\bf r},
t) |0\rangle $. In Eqs. (14) and (15) we shall make replacement of $t ^\prime $ by $t
^ {\prime\prime} =t ^\prime-t $. At averaging $ \langle 0 |... |0 \rangle $ we shall
take into account, that
\begin{equation}
\label {16} \langle
 0|\hat{F}(t)|0\rangle=\langle0|e^{i{\cal{H}}t/\hbar}\hat{F}e^{-i{\cal{H}}t/\hbar}|0\rangle
= \langle0 |\hat {F} |0\rangle,
\end{equation}
where $ \hat {F} $ is any operator. Then we shall obtain
\begin{eqnarray}
\label {17} \langle0|j _ {1\alpha} ({\bf r}, t) |0\rangle =- {e\over mc} \langle0
|\rho ({\bf
r}) |0\rangle A_\alpha ({\bf r}, t) \nonumber\\
+ {i\over\hbar c} \int d^3r^\prime\int_{-\infty}^0dt^\prime\langle0|[j_\alpha ({\bf
r}), j_\beta ({\bf
r} ^ \prime, t ^\prime)] |0\rangle A_\beta ({\bf r} ^ \prime, t+t ^\prime) \nonumber\\
- {i\over \hbar} \int d^3r^\prime\int_{-\infty}^0dt^\prime\langle0|[j_\alpha ({\bf
r}), \rho ({\bf r} ^ \prime, t ^\prime)] |0\rangle\varphi ({\bf r} ^ \prime, t+t
^\prime);
\end{eqnarray}
\begin{eqnarray}
\label {18} \langle0 |\rho_1 ({\bf r}, t)|0\rangle&=&{i\over \hbar c} \int d^3r
^\prime\int _ {-\infty} ^0dt ^\prime \langle0 | [\rho ({\bf r}, t), j_\beta ({\bf
r} ^ \prime, t^\prime)]|0\rangle\nonumber\\&\times& A_\beta ({\bf r} ^ \prime, t+t ^\prime) \nonumber\\
 &-&{i\over \hbar} \int d^3r ^\prime\int _ {-\infty} ^0 dt ^\prime\langle0 | [\rho ({\bf r}),
\rho
( {\bf r} ^ \prime, t ^\prime)] |0\rangle\nonumber\\
 &\times&\varphi({\bf r} ^ \prime, t+t ^\prime).
\end{eqnarray}
Thus, we have obtained expressions for current and charge densities averaged on the
ground state through vector and scalar potentials. But averaged values  should be
expressed through measured values - the fields $ {\bf E} ({\bf r}, t) $ and $ {\bf H}
({\bf r}, t) $ and their derivative. Let us make transition to expressions of $
\langle0 | {\bf j} _ {1} ({\bf r}, t) |0\rangle $ and $ \langle0 |\rho_1 ({\bf r}, t)
|0\rangle $ through fields.

\section{Time derivatives from averaged current and charge densities}
Let us apply the following reception: let us calculate time derivatives from values
(17) and (18):
\begin{eqnarray}
\label {19} {\partial\over \partial t} \langle0|j _ {1\alpha} ({\bf r}, t)|0\rangle&=&
{ \partial\over \partial t} \langle0|j _ {1\alpha} ({\bf r}, t) |0\rangle_A\nonumber\\
 &+&{\partial\over
\partial t} \langle0|j _ {1\alpha} ({\bf r}, t) |0\rangle_\varphi, \nonumber\\{ \partial\over \partial t} \langle0 |\rho_1 ({\bf r}, t)|0\rangle&=&{\partial\over
\partial t} \langle0 |\rho_1 ({\bf r}, t) |0\rangle_A\nonumber\\
 &+&{\partial\over
\partial t} \langle0 |\rho_1 ({\bf r}, t) |0\rangle_\varphi,
\end{eqnarray}
where indexes $A $ and $ \varphi $ designate contributions of vector and scalar
potentials, respectively, equal
\begin{eqnarray}
\label {20} {\partial\over \partial t} \langle0|j _ {1\alpha} ({\bf r}, t) |0\rangle_A
= - {e\over mc} \langle0 |\rho ({\bf
r}) |0\rangle {\partial A_\alpha ({\bf r}, t) \over\partial t} \nonumber\\
+ {i\over\hbar c} \int d^3r^\prime\int_{-\infty}^0dt^\prime\langle0|[j_\alpha ({\bf
r}), j_\beta
( {\bf r} ^ \prime, t ^\prime)] |0\rangle \nonumber\\
\times {\partial A_\beta ({\bf r} ^ \prime, t+t ^\prime) \over \partial t},
\end{eqnarray}
\begin{eqnarray}
\label {21} {\partial\over \partial t} \langle0 |\rho_1 ({\bf r},
t)|0\rangle_A={i\over \hbar c} \int
d^3r^\prime\int_{-\infty}^0dt^\prime\nonumber\\
 \times\langle0|[\rho ({\bf r}), j_\beta ( {\bf
r} ^ \prime, t ^\prime)] |0\rangle{\partial A_\beta ({\bf r} ^ \prime, t+t ^\prime)
\over \partial t},
\end{eqnarray}
\begin{eqnarray}
\label {22} {\partial\over \partial t} \langle0|j _ {1\alpha} ({\bf r}, t)
|0\rangle_\varphi =- {i\over \hbar} \int
 d^3r^\prime\int_{-\infty}^0dt^\prime\nonumber\\
\times\langle0 | [j_\alpha ({\bf r}, t), \rho ({\bf r} ^ \prime, t ^\prime)]
|0\rangle {\partial\varphi ({\bf r} ^ \prime, t+t ^\prime) \over\partial t},
\end{eqnarray}
\begin{eqnarray}
\label {23} {\partial\over \partial t} \langle 0 |\rho_1 ({\bf r}, t)
|0\rangle_\varphi
 =-{i\over \hbar} \int d^3r^\prime\int_{-\infty}^0dt^\prime\nonumber\\
 \times\langle 0 | [\rho ({\bf r}),
\rho ({\bf r} ^ \prime, t ^\prime)] |0\rangle{\partial\varphi({\bf r} ^ \prime, t+t
^\prime) \over
\partial t}.
\end{eqnarray}
Let us transform Eqs. (22) and (23), containing scalar potential $ \varphi $. We use
identity
$$ {\partial\varphi ({\bf r}, t + t ^\prime) \over\partial t} =
{ \partial\varphi ({\bf r}, t + t ^\prime) \over\partial t ^\prime}, $$ then we
integrate on $t ^\prime $ in parts. We obtain
\begin{eqnarray}
\label {24} {\partial\over \partial t} \langle0|j _ {1\alpha} ({\bf r}, t)
|0\rangle_\varphi =- {i\over \hbar} \int d^3r ^\prime\langle 0 | [j_\alpha ({\bf r}),
\rho ({\bf r} ^ \prime)] |0\rangle\nonumber\\
\times\varphi ({\bf r} ^ \prime, t) \nonumber\\
+ {i\over \hbar} \int d^3r^\prime\int_{-\infty}^0dt^\prime\left\langle 0\left |\left
[j_\alpha ({\bf r}) {, \partial\rho ({\bf r} ^ \prime, t ^\prime) \over\partial
 t^\prime}\right]\right|0\right\rangle\nonumber\\
\times\varphi ({\bf r} ^ \prime, t + t ^\prime).
\end{eqnarray}
In the first term of the RHS Eq. (24) we execute integration on $ {\bf r} ^ \prime $
and we use Eq. (6) for operators $j ({\bf r}) $ and $ \rho ({\bf r}) $. To calculate
the second term we use the continuity equation
\begin{equation}
\label {25} \nabla {\bf j} ({\bf r}, t) + {\partial\rho ({\bf r}, t) \over\partial t}
=0,
\end{equation}
which is true for operators, determined in Eq. (10) with taking into account the
constant magnetic field $ {\bf H} _c $. Further in this term we make integration on
$r_\beta ^\prime $ in parts, transferring derivative on the scalar potential $ \varphi
({\bf r} ^ \prime, t+t ^\prime) $. It results in
\begin{eqnarray}
\label {26} {\partial\over \partial t} \langle 0|j _ {1\alpha} ({\bf r}, t)
|0\rangle_\varphi =- {e\over m} \langle 0 |\rho ({\bf r}) |0\rangle {\partial\varphi
({\bf r}, t) \over\partial r_\alpha} \nonumber\\ + {i\over \hbar} \int
 d^3r^\prime\int_{-\infty}^0dt^\prime\langle 0 | [j_\alpha ({\bf r}),
j_\beta ({\bf r} ^ \prime, t ^\prime)] |0\rangle\nonumber\\
\times {\partial\varphi ({\bf r} ^ \prime, t+t ^\prime) \over\partial r _ {\beta} ^
\prime}.
\end{eqnarray}
Summing Eqs. (20) and (26) and using Eq. (1), we obtain
\begin{eqnarray}
\label {27} {\partial\over \partial t} \langle 0|j _ {1\alpha} ({\bf r}, t) |0\rangle
= {e\over m} \langle 0 |\rho ({\bf r}) |0\rangle E_\alpha ({\bf r}, t ) \nonumber\\
- {i\over \hbar} \int d^3r^\prime\int_{-\infty}^0dt^\prime\langle 0 | [j_\alpha ({\bf
r}),
j_\beta ({\bf r} ^ \prime, t ^\prime)] |0\rangle\nonumber\\
\times E_\alpha ({\bf r} ^ \prime, t+t ^\prime).
\end{eqnarray}
We obtain completely similarly
\begin{eqnarray}
\label {28} {\partial\langle0 |\rho_1 ({\bf r}, t) |0\rangle\over \partial
t} = - {i\over \hbar} \int d^3r^\prime\int_{-\infty}^0dt^\prime\nonumber\\
\langle0 | [\rho ({\bf r}), j_\beta ({\bf r} ^ \prime, t ^\prime)] |0\rangle E_\beta
({\bf r} ^ \prime, t+t ^\prime).
\end{eqnarray}
So, we managed to express time derivatives from averaged  current and charge
densities through electric fields, having got rid of vector and scalar potentials.

\section{Average values  of current and
charge densities expressed through electric field}

Integrating Eqs.(27) and (28) on time, we obtain expressions for average current and
charge densities
\begin{eqnarray}
\label {29} \langle0|j _ {1\alpha} ({\bf r}, t) |0\rangle &=&
 \int_{-\infty}^tdt^\prime{\partial\langle 0 | [j _ {1, \alpha} ({\bf r},
t ^\prime) |0\rangle\over
\partial t ^\prime} +C_\alpha, \nonumber\\
\langle 0 |\rho_1 ({\bf r}, t)|0\rangle&=& \int_{-\infty}^tdt^\prime{\partial\langle
0 |\rho_1 ({\bf r}, t ^\prime) |0\rangle\over\partial t ^\prime} +C ^\prime.
\end{eqnarray}
We believe $C_\alpha=C ^\prime=0 $, what corresponds to absence of induced currents
and charges on times, indefinitely removed in the past.

Let us enter a designation
\begin{equation}
\label {30} {\bf a} ({\bf r}, t) = -c\int _ {-\infty} ^tdt ^\prime {\bf E} ({\bf r},
t ^\prime).
\end{equation}
Then with the help Eqs. (27) - (29) we obtain
\begin{eqnarray}
\label {31} \langle0|j _ {1\alpha} ({\bf r}, t) |0\rangle = - {e\over mc} \langle0
|\rho ({\bf
r}, t) |0\rangle a_\alpha ({\bf r}, t) \nonumber\\
+ {i\over\hbar c} \int d^3r^\prime\int_{-\infty}^tdt^\prime\langle0|[j_\alpha ({\bf
r}, t), j_\beta ({\bf r} ^ \prime, t ^\prime)] |0\rangle a_\beta ({\bf r} ^ \prime, t
^\prime),
\end{eqnarray}
\begin{eqnarray}
\label {32} \langle0 |\rho_1 ({\bf r}, t) |0\rangle = \nonumber\\{ i\over \hbar c}
\int d^3r^\prime\int_{-\infty}^tdt^\prime\langle0|[\rho ({\bf r}, t), j_\beta ({\bf
r} ^ \prime, t ^\prime)] |0\rangle a_\beta ({\bf r} ^ \prime, t ^\prime).
\end{eqnarray}

Comparing expressions Eqs. (31) and (32) with Eqs. (17) and (18), we see, that the
second differ from the first by absence of the scalar potential $ \varphi $ and
replacement of the vector potential $ {\bf A} ({\bf r}, t) $ by the vector $ {\bf a}
({\bf r}, t) $, determined in Eq. (30). Thus, we have managed to express average
values of the induced  current and charge densities only through electric fields.
However, without the further transformations Eqs. (31) and (32) are inapplicable at
transition to electric field, independent on time, since at integration on $t $ in Eq.
(30) there appears some uncertainty, - the frequency $ \omega $ in the denominator
turns in zero. It concerns and to the case $ {\bf E} =0, {\bf H} \neq 0 $. We shall
return to this question in section XII.

\section{Transformation of expressions for average
 current and charge densities}
After we managed to express average current and charge densities through an electric
field, the task is to pass to expressions, obviously appropriate to electron-field
interaction of a kind
\begin{equation}
\label {33} {\tilde{U}}_1=-e\sum_ir_{i\beta}E_\beta(t),
\end{equation}
which, for example, is used in \cite{8} for electric field, $ {\bf E} (t) $,
independent on coordinates.

To achieve this purpose we shall make some transformations of Eqs. (31) and (32). We
shall enter fictitious operators $ {\bf j} ^f_1 ({\bf r}, t) $ and $ \rho^f_1 ({\bf
r}, t) $, for which we have
\begin{eqnarray}
\label {34} \langle0|j _ {1\alpha} ({\bf r}, t) |0\rangle = \langle0|j _ {1\alpha} ^f
({\bf r},
t) |0\rangle, \nonumber\\
\langle0 |\rho_1 ({\bf r}, t) |0\rangle =\langle0 |\rho_1^f ({\bf r}, t) |0\rangle.
\end{eqnarray}
It follows from Eqs. (31), (32)
\begin{eqnarray}
\label {35} j _ {1\alpha} ^f ({\bf r}, t) = - {e\over mc} \rho ({\bf
r}, t) a_\alpha ({\bf r}, t) \nonumber\\
+ {i\over\hbar c} \int d^3r^\prime\int_{-\infty}^tdt^\prime[j_\alpha ({\bf r}, t),
j_\beta ({\bf r} ^ \prime, t ^\prime)] a_\beta ({\bf r} ^ \prime, t ^\prime),
\end{eqnarray}
\begin{eqnarray}
\label {36} \rho_1^f ({\bf r}, t)
= \nonumber\\
= {i\over \hbar c} \int d^3r^\prime\int_{-\infty}^tdt^\prime[\rho ({\bf r}, t),
j_\beta ( {\bf r} ^ \prime, t ^\prime)] a_\beta ({\bf r} ^ \prime, t ^\prime).
\end{eqnarray}
Comparing the fictitious operators Eqs. (35) - (36) with the genuine operators Eqs.
(14) - (15), we find, that for transition from genuine  to the fictitious operators
it is necessary to put $ \varphi ({\bf r}, t) =0 $ and to replace the vector potential
$ {\bf A} ({\bf r}, t) $ by  $ {\bf a} ({\bf r}, t) $, determined in Eq. (30).

Let us enter also a fictitious operator of interaction of particles with a field
\begin{equation}
\label {37} U_1^f =- {1\over c} \int d^3r {\bf j} ({\bf r}) {\bf a} ({\bf r}, t),
\end{equation}
which differs from the true operator $U_1 $, determined in Eq. (5) by condition $
\varphi ({\bf r}, t) =0 $ and by replacement $ {\bf A} ({\bf r}, t) $ on $ {\bf a}
({\bf r}, t) $ .To the interaction Eq. (37) there corresponds a linear on a field the
additive to the current density operator
\begin{equation}
\label {38} \Delta j_\alpha^f =- {e\over mc} \rho ({\bf r}) a_\alpha ({\bf r}, t).
\end{equation}
It is easy to see, that Eqs. (35) and (36) may be written as
\begin{equation}
\label {39} j _ {1\alpha} ^f ({\bf r}, t) = \Delta j _ {\alpha} ^f ({\bf r}, t) -
{i\over\hbar}\int_{-\infty}^tdt^\prime[j_\alpha ({\bf r}, t), U^f_1 (t ^\prime)],
\end{equation}
\begin{equation}
\label {40} \rho_1^f ({\bf r}, t) = - {i\over \hbar} \int _ {-\infty} ^tdt ^\prime
[\rho ({\bf r}, t), U_1^f (t ^\prime)],
\end{equation}
what is similar to Eqs. (13a) and (13b) with replacement of the genuine operators by
fictitious. Let us transform Eqs. (39) - (40) so that to remove the first term in the
RHS Eq. (39). It is possible to rewrite  the integral of a kind
$$\int _ {-\infty} ^tdt ^\prime [F ({\bf r}, t), U_1^f (t ^\prime)] $$
from the RHSs Eqs. (39) and (40) as
\begin{equation}
\label {41} \int _ {-\infty} ^tdt ^\prime [F ({\bf r}, t), \tilde {U} _1 (t ^\prime)]
- [ F ({\bf r}, t), R (t)],
\end{equation}
if
\begin{equation}
\label {42} {\tilde U} _1=U_1^f (t) + {dR (t) \over dt},
\end{equation}
where $F ({\bf r}, t) $ is the operator, in case Eq. (39) equal $j_\alpha ({\bf r}, t)
$, and in case Eq. (40) - $ \rho ({\bf r}, t) $, and $R (t) $ is any operator in the
interaction representation
\begin{equation}
\label {43} R (t) =e^{i {\cal {H}} t/\hbar} R_{Sch}({\bf r})
e^{-i {\cal {H}} t/\hbar},
\end{equation}
$R_{Sch} $ is the operator in the Schr\"odinger representation. It is possible to
show, that if
\begin{equation}
\label {44} R _ {Sch} =R _ {Sch} ({\bf r} _1\ldots {\bf r} _N, t),
\end{equation}
i.e. the operator $R _ {Sch} $ does not contain momentums, the relation is carried out
\begin{equation}
\label {45} {i\over\hbar} [j_\alpha ({\bf r}, t), R (t)] = \Delta j _ {R\alpha} ({\bf
r}, t),
\end{equation}
where
\begin{eqnarray}
\label {46} \Delta j _ {R\alpha} ({\bf r}, t)&=&{e\over 2} e ^ {i {\cal {H}} t/\hbar}
\sum_i [\Delta
v _ {iR\alpha} \delta ({\bf r} - {\bf r} _i) \nonumber\\
 &+&\delta({\bf r} - {\bf r} _i) \Delta v _ {iR\alpha}] e ^ {-i {\cal {H}} t/\hbar},
\end{eqnarray}
\begin{equation}
\label {47} \Delta v _ {iR\alpha} = {i\over\hbar} [\dot {R} _ {Sch}, r _ {i\alpha}],
\end{equation}
\begin{equation}
\label {48} \dot {R} _ {Sch} = {i\over\hbar} [{\cal {H}}, R _ {Sch}] + {\partial R _
{Sch} \over\partial t}.
\end{equation}
Also it is obvious, that under condition of Eq. (44),
\begin{equation}
\label {49} [\rho ({\bf r}, t), R (t)] =e ^ {i {\cal {H}} t/\hbar} [\rho ({\bf r}), R
_ {Sch}] e ^ {-i {\cal {H}} t/\hbar} =0.
\end{equation}
Thus, it is proved, that instead of $U_1^f (t) $ it is possible to choose any
operator, determined in Eq. (42), if $R _ {Sch} $ does not contain momentums, and
instead of Eqs. (39) and (40) to write down
\begin{equation}
\label {50} j _ {1\alpha} ^f ({\bf r}, t) = \Delta {\tilde j} _ {\alpha} ({\bf r}, t)
- {i\over\hbar}\int_{-\infty}^tdt^\prime[j_\alpha ({\bf r}, t), \tilde {U} _1 (t
^\prime)],
\end{equation}
\begin{equation}
\label {51} \rho_1^f ({\bf r}, t) = - {i\over \hbar} \int _ {-\infty} ^tdt ^\prime
[\rho ({\bf r}, t), \tilde {U} _1 (t ^\prime)],
\end{equation}
where
\begin{equation}
\label {52} \Delta \tilde {j} _ {\alpha} ({\bf r}) = {e\over 2} \sum_i [\Delta \tilde
{v} _ {i\alpha} \delta ({\bf r} - {\bf r} _i) + \delta ({\bf r} - {\bf r} _i) \Delta
\tilde {v} _ {i\alpha}],
\end{equation}
\begin{equation}
\label {53} \Delta \tilde{v}_{i\alpha}={i\over\hbar}[\tilde{U}_1, r _ {i\alpha}].
\end{equation}
Having substituted Eqs. (42) and (52) in Eqs. (50) and (51), we obtain
\begin{eqnarray}
\label {54} j _ {1\alpha} ^f ({\bf r}, t) = - {e\over mc} \rho ({\bf r}, t) a_\alpha
({\bf r}, t) + {i\over\hbar} [j _ {\alpha} ({\bf r}, t)
, R (t)] \nonumber\\
+ {i\over\hbar c} \int d^3r^\prime\int_{-\infty}^tdt^\prime[j_\alpha ({\bf r}, t),
j_\beta ({\bf r} ^ \prime, t ^\prime)] a_\beta ({\bf r} ^ \prime,
t ^\prime) \nonumber\\
- {i\over\hbar}\int^t_{-\infty}dt^\prime\left[j_\alpha({\bf r}, t), {dR (t ^\prime)
\over dt ^\prime} \right],
\end{eqnarray}
\begin{eqnarray}
\label {55} \rho _ {1\alpha} ^f ({\bf r}, t) = \nonumber\\{ i\over \hbar c} \int
d^3r^\prime\int_{-\infty}^tdt^\prime[\rho ({\bf r}, t), j_\beta ( {\bf r} ^ \prime, t
^\prime)] a_\beta ({\bf r} ^ \prime,
t ^\prime) \nonumber\\
- {i\over\hbar}\int^t_{-\infty}dt^\prime\left[\rho({\bf r}, t), {dR (t ^\prime) \over
dt ^\prime} \right].
\end{eqnarray}

\section{Choice of the operator $R_{Sch} $}
Let us choose the operator $R _ {Sch} $ as
\begin{equation}
\label {56} R _ {Sch} = {e\over c} \sum_i r _ {i\beta} a_\beta ({\bf r} _i, t).
\end{equation}
Then
\begin{eqnarray}
\label {57} R (t) =e ^ {i {\cal {H}} t/\hbar} R _ {Sch} ({\bf r}) e ^ {-i {\cal {H}}
t/\hbar}\nonumber\\ = {1\over c} \int d^3r d_\beta ({\bf r}, t) a_\beta ({\bf r}, t),
\end{eqnarray}
where the designation is entered
\begin{equation}
\label {58} {\bf d} ({\bf r}) =e\sum_i {\bf r} _i\delta ({\bf r} - {\bf r} _i) =e {\bf
r} \sum_i\delta ({\bf r} - {\bf r} _i) = {\bf r} \rho ({\bf r}).
\end{equation}
Let us calculate the contributions in the RHSs Eqs. (54) and (55), containing $R (t) $
and $dR (t) /dt $. It is possible to show, that
\begin{eqnarray}
\label {59} {i\over \hbar} [j_\alpha ({\bf r}, t), R (t)] = {e\over mc} \rho ({\bf r},
t) a_\alpha ({\bf r}, t)\nonumber\\ + {e\over mc} d_\beta ({\bf r}, t) {\partial
a_\beta ({\bf r}, t) \over\partial r_\alpha}.
\end{eqnarray}
Also it is possible to obtain the result
\begin{eqnarray}
\label {60} {dR (t) \over dt} = {1\over c} \int d^3rd_\beta ({\bf r},
t) {\partial a_\beta ({\bf r}, t) \over \partial t} \nonumber\\
+ {1\over c} \int d^3rj_\beta ({\bf r}, t) a_\beta ({\bf r}, t) \nonumber\\
+ {1\over c} \int d^3rY _ {\beta\gamma} ({\bf r}, t) {\partial a_\beta ({\bf r}, t)
\over\partial r_\gamma},
\end{eqnarray}
where
\begin{equation}
\label {61} Y _ {\beta\gamma} ({\bf r}) =r_\beta j_\gamma ({\bf r}).
\end{equation}
Substituting Eqs. (59) and (60) in the RHS Eqs. (54) and (55), we find, that the
terms, not containing derivative $ \partial a_\beta ({\bf r}, t) /\partial t $ or $
\partial a_\beta ({\bf r}, t) /\partial r_\alpha $, are reduced. In a result we have
\begin{eqnarray}
\label {62} j^f _ {1\alpha} = {e\over mc} d_\beta ({\bf r}, t) {\partial a_\beta ({\bf
r}, t) \over \partial r_\alpha} \nonumber\\
- {i\over\hbar c} \int d^3r^\prime\int_{-\infty}^tdt^\prime[j_\alpha({\bf r}, t),
d_\beta ({\bf r} ^ \prime, t ^\prime)] {\partial a_\beta ({\bf r} ^ \prime, t
^\prime) \over
\partial
t ^\prime} \nonumber\\
- {i\over\hbar c} \int d^3r^\prime\int_{-\infty}^tdt^\prime[j_\alpha({\bf r}, t), Y _
{\beta\gamma} ({\bf r} ^ \prime, t ^\prime)] {\partial a_\beta ({\bf r} ^ \prime, t
^\prime) \over \partial r ^\prime_\gamma},
\end{eqnarray}
\begin{eqnarray}
\label {63} \rho^f_1 ({\bf r}, t) = \nonumber\\
= - {i\over\hbar c} \int
 d^3r^\prime\int_{-\infty}^tdt^\prime[\rho({\bf r}, t), d_\beta ({\bf r} ^ \prime,
t ^\prime)] {\partial a_\beta ({\bf r} ^ \prime, t ^\prime) \over \partial
t ^\prime} \nonumber\\
- {i\over\hbar c} \int d^3r^\prime\int_{-\infty}^tdt^\prime[\rho({\bf r}, t), Y _
{\beta\gamma} ({\bf r} ^ \prime, t ^\prime)] {\partial a_\beta ({\bf r} ^ \prime, t
^\prime) \over \partial r ^\prime_\gamma}.
\end{eqnarray}
Averaging the fictitious operators Eqs. (62) and (63) on the ground state, we shall
obtain required by us expressions for averaged values of the induced current and
charge densities.

Taking into account the  definition Eq. (30), we find, that terms in the RHS Eqs. (62)
and (63) are divided on two categories:  to the first concern what contain an
electric field, to the  second - what contain derivative from this field on
coordinates. Therefore it is convenient to write down average values from Eqs. (62)
and (63) as
\begin{equation}
\label {64} \langle 0|j _ {1\alpha} ({\bf r}, t) |0\rangle =\langle 0|j _ {1\alpha}
({\bf r}, t) |0\rangle_E +\langle 0|j _ {1\alpha} ({\bf r}, t) |0\rangle _ {\partial
E/\partial r ^\prime},
\end{equation}
\begin{equation}
\label {65} \langle 0 |\rho _ {1} ({\bf r}, t) |0\rangle =\langle 0 |\rho _ {1} ({\bf
r}, t) |0\rangle_E +\langle 0 |\rho _ {1} ({\bf r}, t) |0\rangle _ {\partial
E/\partial r},
\end{equation}
where
\begin{eqnarray}
\label {66} \langle 0|j _ {1\alpha} ({\bf r}, t) |0\rangle_E
=\nonumber\\{i\over\hbar} \int d^3r ^\prime \int _ {-\infty} ^tdt ^\prime\langle 0 |
[j_\alpha ({\bf r}, t), d_\beta ({\bf r} ^ \prime, t ^\prime)] |0\rangle E_\beta
({\bf r} ^ \prime, t ^\prime),
\end{eqnarray}
\begin{eqnarray}
\label {67} \langle 0 |\rho _ {1} ({\bf r}, t) |0\rangle_E = {i\over\hbar} \int
d^3r ^\prime\nonumber\\
\times\int _ {-\infty} ^tdt ^\prime\langle 0 | [\rho ({\bf r}, t), d_\beta ({\bf r} ^
\prime, t ^\prime)] |0\rangle E_\beta ({\bf r} ^ \prime, t ^\prime),
\end{eqnarray}
\begin{eqnarray}
\label {68} &\langle& 0|j _ {1\alpha} ({\bf r}, t) |0\rangle _ {\partial E/\partial
r} = - {e\over m} \langle 0|d_\beta ({\bf
r})|0\rangle\int_{-\infty}^tdt^\prime{\partial E_\beta ({\bf
r}, t ^\prime) \over\partial r_\alpha} \nonumber\\
 &+&{i\over\hbar}\int d^3r^\prime\int_{-\infty}^tdt^\prime\langle 0 | [j_\alpha ({\bf r},
t), Y _ {\beta\gamma} ({\bf r} ^ \prime,
t ^\prime)] |0\rangle\nonumber\\
 &\times&\int_{-\infty}^{t^\prime}dt^{\prime\prime}{\partial E_\beta ({\bf r} ^ \prime,
t ^ {\prime\prime}) \over\partial r ^\prime_\gamma},
\end{eqnarray}
\begin{eqnarray}
\label {69} \langle 0 |\rho_1 ({\bf r}, t) |0\rangle _ {\partial E/\partial r} =
{i\over\hbar} \int d^3r^\prime\int_{-\infty}^tdt^\prime\nonumber\\
\times\langle0 | [\rho ({\bf r}, t), Y _ {\alpha\beta} ({\bf r} ^ \prime,
 t^{\prime})]|0\rangle\int_{-\infty}^{t^\prime}dt^{\prime\prime}{\partial E_\beta ({\bf
r} ^ \prime, t ^ {\prime\prime}) \over\partial r ^\prime_\gamma}.
\end{eqnarray}
So, the task of expression of averaged on the ground state current and charge
densities in linear approximation on electric field and its derivatives on
coordinates is solved.

\section{The transformed fictitious operator of interaction}
For completeness of a picture we shall determine a kind of the fictitious operator of
interaction $ \tilde {U} _1=U_1^f +\dot {R} _ {Sch} $. Using Eqs. (37) and (60) we
obtain in the Schr\"odinger representation
\begin{equation}
\label {70} \tilde {U} _1 = \tilde {U} _ {1E} + \tilde {U} _ {1\partial E/\partial r},
\end{equation}
\begin{equation}
\label {71} \tilde {U} _ {1E} = -\int d^3r d_\beta ({\bf r}) E_\beta ({\bf r}, t),
\end{equation}
\begin{equation}
\label {72} \tilde {U} _ {1\partial E/\partial r} = -\int d^3r Y _ {\beta\gamma}
({\bf r}) \int _ {-\infty} ^t dt ^\prime {\partial E_\beta ({\bf r}, t ^\prime)
\over\partial r_\gamma}.
\end{equation}
Having executed integration on $ {\bf r} $ in the RHS Eqs. (71) and (72), we obtain
\begin{equation}
\label {73} \tilde{U}_{1E}=-e\sum_ir_{i\beta}E_\beta({\bf r} _i, t),
\end{equation}
\begin{eqnarray}
\label {74} \tilde {U}_{1\partial E/\partial r} = - {e\over 4} \int_{-\infty}^t dt
^\prime\sum_i \left \{\left (v_{i\gamma}{\partial E_\beta ({\bf r}_i,
t^\prime) \over\partial r_{i\gamma}} \right.\right.\nonumber\\
+ \left. {\partial E_\beta ({\bf r} _i, t ^\prime) \over\partial
 r_{i\gamma}}v_{i\gamma}\right)r_{i\beta}\nonumber\\
 +\left.r_{i\beta}\left(v_{i\gamma}{\partial E_\beta ({\bf r} _i, t ^\prime) \over\partial
r _ {i\gamma}} + {\partial E_\beta ({\bf r} _i, t ^\prime) \over\partial
 r_{i\gamma}}v_{i\gamma}\right)\right\}\nonumber\\
= -{e\over 2} \int_{-\infty}^t dt ^\prime\sum_i \left \{v _ {i\gamma} {\partial
E_\beta ({\bf r} _i, t ^\prime) \over\partial
r_{i\gamma}} r_{i\beta} \right.\nonumber\\
+ \left.r_{i\beta} {\partial E_\beta ({\bf r}_i, t^\prime) \over\partial r_
{i\gamma}} v _ {i\gamma} \right \}.
\end{eqnarray}
Thus, $ \tilde {U} _ {1E} $ contains an electric field, and $
\tilde {U} _ {1\partial E/\partial r} $ contains derivatives from
electric field on  coordinates and an integral on time. In the
case, when terms, containing derivatives from electric field on
coordinates, on what of reasons give the small contribution
\footnote { Strictly speaking, it follows from the Maxwell
equations, that if the electric field depends on time, it also
depends on coordinate, i. e. derivatives  of field components on
coordinates are distinct from zero.}, it is possible to use Eq.
(73). It coincides with the formula for interaction of charged
particles in \cite{8,9,10}. Let us notice, that Eq. (70) for
fictitious interaction $ \tilde {U} _1 $ may be written  compactly
\begin{equation}
\label {75} \tilde {U} _1 = {e\over 2c} \sum_i\left \{{da_\beta ({\bf r} _i, t) \over
dt} r _ {i\beta} +r _ {i\beta} {da_\beta ({\bf r} _i, t) \over dt} \right \},
\end{equation}
where $da_\beta ({\bf r} _i, t) /dt $ is the full time derivative:
\begin{equation}
\label {76} {da_\beta ({\bf r} _i, t) \over dt} = {\partial a_\beta ({\bf r} _i, t)
\over
\partial t} + {i\over\hbar} [{\cal {H}}, a_\beta ({\bf r} _i, t)],
\end{equation}
and
\begin{eqnarray}
\label {77} {i\over\hbar} [{\cal {H}}, a_\beta ({\bf r} _i, t)] = {1\over 2} \left
\{v _ {i\gamma} {\partial a_\beta ({\bf r} _i, t) \over\partial
r _ {i\gamma}} \right.\nonumber\\
+ \left. {\partial a_\beta ({\bf r} _i, t) \over\partial r _ {i\gamma}} v _ {i\gamma}
\right \}.
\end{eqnarray}

The  following linear on a field  additive to speed corresponds to interaction Eq.
(70)
\begin{equation}
\label {78} \Delta\tilde {v} _ {i\alpha} = {e\over mc} r _ {i\beta} {\partial a_\beta
({\bf r} _i, t) \over\partial r _ {i\alpha}},
\end{equation}
what corresponds, according to Eq. (52), to an  additive to the current density
\begin{equation}
\label {79} \Delta\tilde {j} _ \alpha ({\bf r}) = {e\over mc} d_\beta ({\bf r})
{\partial a_\beta ({\bf r}, t) \over\partial r _ {\alpha}}.
\end{equation}
Passing to the interaction representation, we obtain the first term in the RHS Eq.
(62).

\section{Exception of diagonal elements of operators of particle coordinates}

Let us return to Eqs. (64) - (69) for average induced  current
and charge densities at $T=0 $, obtained  in section VIII. Taking
into account definition (58) of operators ${\bf d}({\bf r}) $,
and also definition (61) of operators $Y_{\beta\gamma} ({\bf
r})$, which may be rewritten as
\begin{equation}
\label{80}Y_{\beta\gamma}({\bf r})={1\over
2}\sum_i(j_{i\gamma}r_{i\beta}+r_{i\beta}j_{i\gamma}),
\end{equation}
we see, that these operators contain coordinates ${\bf r}_i$ of
particles. But averaged values $\langle 0| {\bf j}_1({\bf r},
t)|0\rangle$ and $\langle0|\rho_1({\bf r}, t) |0\rangle$ do not
owe to depend on an initial  point of readout of coordinates $
{\bf r}_i $. Let's transform expressions (64) - (69) so that last
property became obvious. Let us represent a vector $ {\bf r} _i $
as two parts
\begin{equation}
\label{81}{\bf r}_i={\bar{\bf r}}_i+\langle 0|{\bf r}_i |0\rangle.
\end{equation}
It is obvious, that matrix elements of the operator ${\bar {\bf
r}}_i$, as diagonal, and not diagonal, do not vary at change
points of readout of coordinates ${\bf r}_i$. We shall show, that
in expressions for average values $\langle0|j_{1\alpha}({\bf r},
t)|0\rangle$ and $\langle0|\rho_1({\bf r}, t)|0\rangle$ the
operators ${\bf r}_i$ may be replaced by ${\bar {\bf r}}_i$. Let's
write down average $\langle0|j{_1\alpha}({\bf r}, t) |0\rangle$ as
\begin{equation}
\label{82}\langle0|j_{1\alpha}({\bf r}, t )|0\rangle=
\overline{\langle0|j_{1\alpha}({\bf r}, t )|0\rangle}
+\sum_i\langle 0| r_{i\beta} |0\rangle x_{i\alpha\beta}({\bf r},
t ),
\end{equation}
where $\overline{\langle0|j_{1\alpha}({\bf r}, t)|0\rangle}$ - is
the contribution of the operators ${\bar {\bf r}}_i$,
\begin{eqnarray}
\label{83}&x&_{i\alpha\beta}({\bf r}, t)={i\over\hbar}\int
d^3r^\prime\int_{-\infty}^tdt^\prime\langle0|[j_\alpha({\bf r},
t),\rho_i({\bf r}^\prime, t^\prime)]|0\rangle\nonumber\\
&\times&E_\beta({\bf r}^\prime, t^\prime)+{e\over
mc}\langle0|\rho_i({\bf r})|0\rangle{\partial a_\beta({\bf r},
t)\over\partial
r_\alpha}\nonumber\\
&-&{i\over\hbar c}\int
d^3r^\prime\int_{-\infty}^tdt^\prime\langle0|[j_\alpha({\bf r},
t),j_{i\gamma}({\bf r}^\prime, t^\prime)]|0\rangle\nonumber\\
&\times&{\partial a_\beta({\bf r}^\prime, t^\prime)\over\partial
r_\gamma^\prime}.
\end{eqnarray}
Having executed in the first and third terms of the RHS Eq. (82) integration on $ {\bf
r} ^ \prime $, we obtain
\begin{eqnarray}
 \label{84}x_{i\alpha\beta}({\bf r}, t) = - {ie\over\hbar c}
 \int_{-\infty}^tdt^\prime\nonumber\\
 \times
 \left\langle0\left|\left[j_\alpha({\bf r}, t), {da_\beta ({\bf
r} _i (t ^\prime), t ^\prime) \over dt ^\prime} \right] \right|0\right\rangle\nonumber\\
+{e\over mc} \langle0 |\rho_i ({\bf r}) |0\rangle {\partial a_\beta ({\bf r}, t)
\over\partial r_\alpha},
\end{eqnarray}
where the designation for the full time derivative is used
\begin{eqnarray}
\label {85} {da_\beta ({\bf r} _i (t), t) \over dt} = {d\over dt}
\left (e ^ {i {\cal {H}} t/\hbar} a_\beta ({\bf
r} _i, t)e^{-i{\cal{H}}t/\hbar}\right)\nonumber\\
=e ^ {i {\cal {H}} t/\hbar} {da_\beta ({\bf r} _i, t) \over dt} e ^ {-i {\cal {H}}
t/\hbar},
\end{eqnarray}
and the derivative $da_\beta ({\bf r} _i, t) /dt $ is determined in Eq. (76). Having
executed in the first term of the RHS Eq. (83) integration on $t ^\prime $ and having
calculated the commutator $ [j_\alpha ({\bf r}), a_\beta ({\bf r} _i, t)], $ we
obtain the result
\begin{equation}
\label {86} x _ {i\alpha\beta} ({\bf r}, t) =0.
\end{equation}
\begin{equation}
\label{87} \langle0|j_{1\alpha}({\bf r}, t )|0\rangle=
\overline{\langle0|j_{1\alpha}({\bf r}, t )|0\rangle}.
\end{equation}

It is similarly possible to obtain
\begin{equation}
\label{88} \langle0|\rho_{1}({\bf r}, t )|0\rangle=
\overline{\langle0|\rho_{1}({\bf r}, t )|0\rangle}.
\end{equation}
Let us represent Eqs. (87) and (88) as two parts
\begin{equation}
\label{89} \langle0|j_{1\alpha}({\bf r}, t )|0\rangle=
\langle0|j_{1\alpha}({\bf r}, t )|0\rangle_I+
\langle0|j_{1\alpha}({\bf r}, t )|0\rangle_{II},
\end{equation}
\begin{equation}
\label{90} \langle0|\rho_{1}({\bf r}, t )|0\rangle=
\langle0|\rho_{1}({\bf r}, t )|0\rangle_I+ \langle0|\rho_{1}({\bf
r}, t )|0\rangle_{II},
\end{equation}
where
\begin{eqnarray}
\label{91}\langle0| j_{1\alpha}({\bf r},
t)|0\rangle_I=\nonumber\\
{i\over \hbar}\int d^3r\prime\int^t_{-\infty}dt^\prime
\langle0|[j_\alpha({\bf r}, t), {\bar d}_\beta({\bf r}^\prime,
t^\prime)]|0\rangle E_\beta({\bf r}^\prime, t^\prime),
\end{eqnarray}
\begin{eqnarray}
\label{92} \langle0| j_{1\alpha}({\bf r},
t)|0\rangle_{II}={e\over mc}\langle 0|{\bar d}_\beta(\bf
r)|0\rangle
{\partial a_\beta({\bf r}, t)\over\partial r_\alpha}\nonumber\\
-{i\over c\hbar}\int d^3r^\prime\int_{-\infty}^tdt^\prime
 \langle 0|[j_\alpha({\bf r},
t){\bar Y}_{\beta\gamma}({\bf r}^\prime,
t^\prime)]|0\rangle\nonumber\\
\times{\partial a_\beta({\bf r}^\prime, t^{\prime})\over\partial
r_\gamma^\prime},
\end{eqnarray}
\begin{eqnarray}
\label{93}\langle0|\rho_{1}({\bf r}, t)|0\rangle_I=\nonumber\\
{i\over \hbar}\int d^3r^\prime\int^t_{-\infty}dt^\prime
\langle0|[\rho({\bf r}, t), {\bar d}_\beta({\bf r}^\prime,
t^\prime)]|0\rangle E_\beta({\bf r}^\prime, t^\prime),
\end{eqnarray}
\begin{eqnarray}
\label{94}\langle 0| \rho_1({\bf r},
t)|0\rangle_{II}=-{i\over\hbar c}\int
d^3r^\prime\int_{-\infty}^tdt^\prime\nonumber\\
\times\langle0|[\rho({\bf r}, t),{\bar Y}_{\beta\gamma}({\bf
r}^\prime, t^\prime)]|0\rangle {\partial a_\beta({\bf r}^\prime,
t^{\prime})\over\partial r_\gamma^\prime},
\end{eqnarray}
and
\begin{eqnarray}
\label{95}{\bar {\bf d}}({\bf r})=e\sum_i{\bar{\bf
r}}_{i}\rho({\bf r}),
\end{eqnarray}
\begin{equation}
\label{96}{\bar Y}_{\beta\gamma}({\bf r})={1\over
2}\sum_i(j_{i\gamma}{\bar r}_{i\beta}+{\bar
r}_{i\beta}j_{i\gamma}).
\end{equation}
{ \it Results (89) - (94) are basic in the present work}. Let's
emphasize, that splittings  Eq. (89) and Eq. (90) average sizes on
two parts do not coincide with splittings Eq. (64) and Eq. (65).

 The contributions with an index $I $ we shall name basic, as they do not
 disappear in case of an electrical field independent from
 coordinates ${\bf r}$. Contributions $II$ contain derivative from
 an electrical field on coordinates.

\section{The conductivity tensor, dependent on coordinates}

Let us consider at first only basic part of the induced  current density, designated
in Eq. (90) by index $I $. We shall write down Eq. (91) as
\begin{eqnarray}
\label {97} \langle0 | j _ {1\alpha} ({\bf r},
t) |0\rangle_I =\nonumber\\
= \int d^3r\int ^\infty _ {-\infty} dt ^\prime l _ {\alpha\beta} ({\bf r}, t, {\bf r}
^ \prime, t ^\prime) E_\beta ({\bf r} ^ \prime, t ^\prime),
\end{eqnarray}
where
\begin{equation}
\label {98} l _ {\alpha\beta} ({\bf r}, t, {\bf r} ^ \prime, t ^\prime) =
{i\over\hbar}
 \Theta(t-t^\prime)\langle0|[j_\alpha({\bf r}, t), \bar {d} _ \beta ({\bf r} ^ \prime,
t ^\prime)] |0\rangle.
\end{equation}
Let us enter a tensor
\begin{equation}
\label {99} \sigma _ {I\alpha\beta} ({\bf r} ^ \prime, t ^\prime | {\bf r}, t) =l _
{\alpha\beta} ({\bf r}, t, {\bf r} - {\bf r} ^ \prime, t-t ^\prime).
\end{equation}
Designation Eq. (99) with a partition we have borrowed from \cite{11}. Then it is
possible to rewrite Eq. (97) as
\begin{eqnarray}
\label {100} \langle0 | j _ {1\alpha} ({\bf r}, t) |0\rangle_I =\nonumber\\ = \int
 d^3r^\prime\int^\infty_{-\infty}dt^\prime\sigma_{I\alpha\beta}({\bf r} ^ \prime,
t ^\prime | {\bf r}, t) E_\beta ({\bf r} - {\bf r} ^ \prime, t-t ^\prime),
\end{eqnarray}
where
\begin{equation}
\label {101} \sigma _ {I\alpha\beta} ({\bf r} ^ \prime, t ^\prime | {\bf r}, t) =
{i\over\hbar} \Theta (t ^\prime) \langle0 | [j_\alpha ({\bf r}, t), \bar {d} _ \beta
({\bf r} - {\bf r} ^ \prime, t-t ^\prime)] |0\rangle.
\end{equation}
It is visible from Eq. (101), that the tensor $ \sigma _ {I\alpha\beta} ({\bf r} ^
\prime, t ^\prime | {\bf r}, t) $  does not depend on $t $!

Now we shall make the Fourier transformation. Let us write down electric field as
\begin{equation}
\label {102} E_\alpha ({\bf r}, t) =E ^ {(+)}_\alpha ({\bf r}, t) +E ^ {(-)}_\alpha
({\bf r}, t),
\end{equation}
where
\begin{equation}
\label {103} E ^ {(+)}_\alpha ({\bf r}, t) = {1\over (2\pi) ^4} \int d^3k\int_0
^\infty d\omega E_\alpha ({\bf k}, \omega) E ^ {i {\bf kr} -i\omega t},
\end{equation}
\begin{equation}
\label {104} E ^ {(-)}_\alpha ({\bf r}, t) = (E ^ {(+)}_\alpha ({\bf r}, t)) ^ *,
\end{equation}
\begin{equation}
\label {105} E_\alpha ({\bf k}, \omega) = \int d^3r\int _ {-\infty} ^ \infty dt
E_\alpha ({\bf r}, t) E ^ {-i {\bf kr} +i\omega t}.
\end{equation}
Let us enter a Fourier-image of the tensor $ \sigma _ {I\alpha\beta} ({\bf r} ^
\prime, t ^\prime | {\bf r}, 0) $ on variables $ {\bf r} ^ \prime, t ^\prime $
\begin{equation}
\label {106} \sigma _ {I\alpha\beta} ({\bf k}, \omega | {\bf r}) = \int d^3r ^\prime
 \int_{-\infty}^\infty\sigma_{I\alpha\beta}({\bf r} ^ \prime,
t ^\prime | {\bf r}, 0) e ^ {i\omega t ^\prime-i {\bf k} {\bf r} ^ \prime}.
\end{equation}
 Then
\begin{equation}
\label {107} \langle0|j _ {1\alpha} ({\bf r}, t) |0\rangle_I =\langle0|j _ {1\alpha}
({\bf r}, t) |0\rangle_I^{(+)}+\langle0|j_{1\alpha}({\bf r}, t) |0\rangle_I ^ {(-)}
\end{equation}
\begin{eqnarray}
\label {108} \langle0|j _ {1\alpha} ({\bf r}, t) |0\rangle_I ^ {(+)}
=\nonumber\\{1\over (2\pi) ^4} \int d^3k\int_0 ^\infty d\omega \sigma _
{I\alpha\beta} ({\bf k}, \omega | {\bf r}) E_\beta ({\bf k}, \omega) e ^ {i {\bf kr}
-i\omega t},
\end{eqnarray}
\begin{eqnarray}
\label {109} \langle0|j _ {1\alpha} ({\bf r},
 t)|0\rangle_I^{(-)}=(\langle0|j_{1\alpha}({\bf r}, t) |0\rangle_I ^ {(+)}) ^*.
\end{eqnarray}
Substituting Eq. (101) in Eq. (106), we obtain
\begin{eqnarray}
\label {110} \sigma _ {I\alpha\beta} ({\bf k}, \omega | {\bf r}) = {i\over\hbar} \int
d^3r ^\prime\int _ {-\infty} ^ \infty \Theta
( t ^\prime) e ^ {-i {\bf k} {\bf r} ^ \prime+i\omega t ^\prime} \nonumber\\
\times\langle0 | [j_\alpha ({\bf r}), \bar {d} _ \beta ({\bf r} - {\bf r} ^ \prime,
-t ^\prime)] |0\rangle.
\end{eqnarray}

 By similar way we find the
contribution in conductivity with an index $II $. Finally we
obtain
\begin{eqnarray}
\label{111}\langle0|j_\alpha({\bf r},
t)|0\rangle^{(+)}=\nonumber\\{1\over (2\pi)^4}\int
d^3k\int_0^\infty d\omega \sigma_{\alpha\beta}({\bf k}, \omega
|{\bf r})E_\beta ({\bf k}, \omega)e^{i{\bf kr}-i\omega t},
\end{eqnarray}
\begin{equation}
\label{112}\sigma_{\alpha\beta}({\bf k}, \omega |{\bf
r})=\sigma_{I\alpha\beta}({\bf k}, \omega |{\bf r})
+\sigma_{II\alpha\beta}({\bf k}, \omega |{\bf r}).
\end{equation}
The value $\sigma_{I\alpha\beta}({\bf k},\omega|{\bf r})$ is
defined in Eq. (110),
\begin{eqnarray}
\label{113}\sigma_{II\alpha\beta}({\bf k}, \omega |{\bf
r})={ek_\alpha\over
m\omega}\langle0|\bar{d}_\beta({\bf r})|0\rangle\nonumber\\
-{ik_\gamma\over m\omega}\int d^3r^\prime\int_{-\infty}^\infty
dt^\prime\Theta
(t^\prime)e^{-i{\bf k}{\bf r}^\prime+i\omega t^\prime}\nonumber\\
\times\langle0|[j_\alpha({\bf r}),{\bar Y}_{\beta\gamma}({\bf
r}-{\bf r}^\prime, -t^\prime)]|0\rangle.
\end{eqnarray}

The conductivity tensor  $\sigma_{\alpha\beta}({\bf k, \omega|{\bf
r}})$ does not depend from coordinates ${\bf r}$ only in case of
spatially - homogeneous system. At consideration of semiconductor
objects of the lowered dimension dependence of the conductivity
tensor from $ {\bf r} $ is rather essential.
 Dependence of the
conductivity tensor  of the low dimensional semiconductor
objects  on $ {\bf r} $ is rather essential.

In all our previous works  (see, for example, \cite{4,12,13,14})
Eq. (91)was used for calculations of an induced density of a
current.

\section{Transition to expressions containing magnetic field}

Till now we left behind frameworks of our consideration a case, when electric field $
{\bf E} $ does not depend on time, in particular the case $ {\bf E} =0,~~ {\bf H}
({\bf r}, t) =const. $ To consider the last case we shall transform expressions for
the average induced current and charge densities obtained at the end of section $VIII
$, having entered in them magnetic field $ {\bf H} ({\bf r}, t) $. For this purpose
each of values $ < 0|j _ {1\alpha} ({\bf r}, t) |0 > _ {\partial E/\partial r} $ and
$ < 0 |\rho_1 ({\bf r}, t) |0 > _ {\partial E/\partial r} $, determined,
respectively, in Eqs. (68) and (69), we shall break on two parts as follows
\footnote{Here top indexes (+) and (-) have no any relation to the same indexes in
section XI.}
\begin{eqnarray}
\label {114} \langle 0|j _ {1\alpha} ({\bf r}, t) |0\rangle _
{\partial E/\partial
r} = \nonumber\\
= \langle 0|j _ {1\alpha} ({\bf r}, t) |0\rangle ^ {(+)} + \langle 0|j _ {1\alpha}
({\bf r}, t) |0\rangle ^ {(-)},
\end{eqnarray}
\begin{eqnarray}
\label {115} \langle 0 |\rho_1 ({\bf r}, t) |0\rangle _ {\partial E/\partial r} = \nonumber\\
= \langle 0 |\rho_1 ({\bf r}, t) |0\rangle ^ {(+)} + \langle 0 |\rho_1 ({\bf r}, t)
|0\rangle ^ {(-)},
\end{eqnarray}
where
\begin{eqnarray}
\label {116} \langle 0|j _ {1\alpha} ({\bf r}, t) |0\rangle ^
{(\pm)} = {e\over 2mc} \langle
0|d_\beta ({\bf r}) |0\rangle\nonumber\\
\times \left ({\partial a_\beta ({\bf r}, t) \over \partial {\bf r} _ \alpha} \pm
{ \partial a_\alpha ({\bf r}, t) \over \partial {\bf r} _ \beta} \right) \nonumber\\
- {i\over 2\hbar c} \int d {\bf r} ^ \prime \int _ {-\infty} ^t dt ^\prime \langle 0
| [j_\alpha ({\bf r}, t), Y _ {\beta \gamma} ({\bf
r ^\prime}, t ^\prime)] |0\rangle\nonumber\\
\times \left ({\partial a_\beta ({\bf r ^\prime}, t ^\prime) \over
\partial {\bf r} _ \gamma ^\prime}
\pm {\partial a_\gamma ({\bf r} ^ \prime, t ^\prime) \over
\partial {\bf r} _ \beta ^\prime} \right),
\end{eqnarray}
\begin{eqnarray}
\label {117} \langle 0 |\rho_1 ({\bf
r},t)|0\rangle^{\pm}&=&-{i\over 2\hbar c} \int d {\bf r} ^ \prime
\int _ {-\infty} ^t dt^\prime\nonumber\\&\times& \langle 0 | [\rho
({\bf r}, t), Y _ {\beta
\gamma} ({\bf r ^\prime}, t ^\prime)] |0\rangle\nonumber\\
 &\times& \left ({\partial a_\beta ({\bf r ^\prime}, t ^\prime) \over
\partial {\bf r} _ \gamma ^\prime}
\pm {\partial a_\gamma ({\bf r} ^ \prime, t ^\prime) \over
\partial {\bf r} _ \beta ^\prime} \right).
\end{eqnarray}
At first we shall consider contributions with index (-). Let us return to vector $
{\bf A} ({\bf r}, t) $ and scalar $ \varphi ({\bf r}, t) $ potentials. Taking into
account definition Eq. (30) for the vector $ {\bf a} ({\bf r}, t) $ and first of the
formulas Eq. (1), we obtain
\begin{equation}
\label {118} { \bf a} ({\bf r}, t) = {\bf A} ({\bf r}, t) +c\int _
{-\infty} ^t dt ^\prime
\partial\varphi ({\bf r} ^ \prime, t ^\prime) /\partial {\bf r}.
\end{equation}
Having substituted Eq. (119) in the RHSs of expressions $ \langle 0|j _ {1\alpha}
({\bf r}, t) |0\rangle ^ {(-)} $ and $ \langle 0 |\rho_1 ({\bf r}, t) |0\rangle ^
{(-)} $, we find, that contributions from scalar potential $ \varphi $ become zero,
and
\begin{eqnarray}
\label {119} &\langle& 0|j _ {1\alpha} ({\bf r}, t) |0\rangle ^
{(-)} = {e\over 2mc} \langle
0|d_\beta ({\bf r}) |0\rangle\nonumber\\
 &\times& \left ({\partial A_\beta ({\bf r}, t) \over \partial {\bf r} _ \alpha}
- {\partial A_\alpha ({\bf r}, t) \over \partial {\bf r} _ \beta} \right) \nonumber\\
 &-& {i\over 2\hbar c} \int d {\bf r} ^ \prime \int _ {-\infty} ^t dt ^\prime \langle
0 | [j_\alpha ({\bf r}, t), Y _ {\beta \, \gamma} ({\bf
r ^\prime}, t ^\prime)] |0\rangle\nonumber\\
 &\times& \left ({\partial A_\beta ({\bf r ^\prime}, t ^\prime) \over
\partial {\bf r} _ \gamma ^\prime}
- {\partial A_\gamma ({\bf r} ^ \prime, t ^\prime) \over
\partial {\bf r} _ \beta ^\prime} \right),
\end{eqnarray}
\begin{eqnarray}
\label {120} \langle 0 |\rho_1 ({\bf
r},t)|0\rangle^{(-)}&=&-{i\over 2\hbar c} \int d {\bf
r} ^ \prime \int _ {-\infty} ^t dt ^\prime\nonumber\\
 &\times& \langle 0 | [\rho ({\bf r}, t), Y _ {\beta \gamma}
( {\bf r ^\prime}, t ^\prime)] |0\rangle\nonumber\\
 &\times& \left ({\partial A_\beta ({\bf r ^\prime}, t ^\prime) \over
\partial {\bf r} _ \gamma ^\prime}
- {\partial A_\gamma ({\bf r} ^ \prime, t ^\prime) \over
\partial {\bf r} _ \beta ^\prime} \right).
\end{eqnarray}
Taking into account, that $ {\bf H} ({\bf r}, t) =rot \, {\bf A} ({\bf r}, t) $
(second equality from Eq. (1)), we obtain easily
\begin{eqnarray}
\label {121} \langle 0|j _ {1\alpha} ({\bf r}, t) |0\rangle ^
{(-)} = - {e\over 2mc} ({\bf H} ({\bf r}, t) \times {\bf r}) _
\alpha
\langle 0 |\rho ({\bf r}) |0\rangle\nonumber\\
+ {i\over 2\hbar c} \int d {\bf r} ^ \prime\int _ {-\infty} ^t dt ^\prime
( {\bf H} ({\bf r} ^ \prime, t ^\prime) \times {\bf r} ^ \prime) _ \beta\nonumber\\
\times\langle 0 | [j_\alpha ({\bf r}, t), j_\beta ({\bf r} ^ \prime, t ^\prime)]
|0\rangle,
\end{eqnarray}
\begin{eqnarray}
\label {122} \langle 0 |\rho_1 ({\bf r}, t) |0\rangle ^ {-} =
{ i\over 2\hbar c} \int d {\bf r} ^ \prime\int _ {-\infty} ^t dt ^\prime\nonumber\\
\times ({\bf H} ({\bf r} ^ \prime, t ^\prime) \times {\bf r} ^ \prime) _ \beta
\langle 0 | [\rho ({\bf r}, t), j_\beta ({\bf r} ^ \prime, t ^\prime)] |0\rangle.
\end{eqnarray}
Thus, the values $ \langle 0|j _ {1\alpha} ({\bf r}, t) |0\rangle ^ {(-)} $ and $
\langle 0 |\rho_1 ({\bf r}, t) |0\rangle ^ {(-)} $ are expressed through magnetic
field.

Now we shall transform Eqs. (117) and (118) for the contributions with an index (+).
We work under the following circuit. Let us consider, for example, Eq. (117). In the
RHS part
\begin{equation}
\label {123} Y _ {\beta\gamma} ({\bf r} ^ \prime) =r_\beta ^\prime
\, j_\gamma ({\bf r} ^ \prime)
\end{equation}
enters;
 for $j_\gamma ({\bf r}) $ it is easy to deduce a ratio
\begin{equation}
\label {124} j_\gamma ({\bf r}) = \dot {d} _ \gamma ({\bf r}) +
\partial Y _ {\gamma \delta} ({\bf r}) /\partial r_\delta.
\end{equation}
Let us substitute Eqs. (124) and (125) in the second term  of the RHS Eq. (117), which
in result breaks up on two contributions - occurring from $ \partial Y _ {\gamma
\delta} ({\bf r} ^ \prime) /\partial r_\delta ^\prime $ and from $ \dot {d} _ \gamma
({\bf r}) $. In the first contribution we integrate in parts on $r_\delta ^\prime $,
in second - on variable $t ^\prime $ also in parts. It results in
\begin{eqnarray}
\label {125} \langle 0|j _ {1\alpha} ({\bf r}, t) |0\rangle ^
{(+)} =-\langle 0|j _ {1\alpha} ({\bf r}, t) |0\rangle ^
{(+)}\nonumber\\- {e\over mc} \langle 0 |\rho ({\bf r}) |0\rangle
r_\beta r_\gamma {\partial^2a_\beta ({\bf r}, t) \over\partial
r_\alpha\partial r_\gamma} \nonumber\\ + {i\over\hbar c} \int d
{\bf r} ^ \prime r_\beta ^\prime r_\gamma ^\prime \int _
{-\infty} ^tdt ^\prime\langle 0 | [j_\alpha ({\bf r}, t),
j_\delta ({\bf r} ^ \prime, t ^\prime)] |0\rangle\nonumber\\\
times {
\partial^2a_\beta ({\bf r} ^ \prime, t ^\prime) \over
\partial r_\gamma ^\prime\partial r_\delta ^\prime}
+ {i\over \hbar c} \int d {\bf r} ^ \prime r_\beta ^\prime
\int _ {-\infty} ^tdt ^\prime\nonumber\\
\times \langle 0 | [j_\alpha ({\bf r}, t), d_\gamma ({\bf r} ^ \prime, t ^\prime)]
|0\rangle { \partial^2a_\beta ({\bf r} ^ \prime, t ^\prime) \over
\partial r_\gamma ^\prime\partial t ^\prime}.
\end{eqnarray}
Using a ratio
$$-c ^ {-1} \partial a_\beta ({\bf r}, t) /\partial t=E_\beta ({\bf r}, t), $$
 we obtain finally from Eq. (125):
\begin{eqnarray}
\label {126} \langle 0|j _ {1\alpha} ({\bf r}, t) |0\rangle ^
{(+)} = - {e\over 2mc} \langle 0 |\rho ({\bf r}) |0\rangle
r_\beta r_\gamma {\partial^2a_\beta ({\bf r}, t) \over
\partial r_\alpha\partial r_\gamma} \nonumber\\ +
{ i\over 2\hbar c} \int d {\bf r} ^ \prime r_\beta ^\prime r_\gamma ^\prime \int _
{-\infty} ^tdt ^\prime\langle 0 | [j_\alpha ({\bf r}, t), j_\delta ({\bf
r} ^ \prime, t ^\prime)] |0\rangle\nonumber\\
\times {\partial^2a_\beta ({\bf r} ^ \prime, t ^\prime) \over
\partial r_\gamma ^\prime\partial r_\delta ^\prime} -
{ i\over 2\hbar} \int d {\bf r} ^ \prime r_\beta ^\prime r_\gamma ^\prime \int _
{-\infty} ^t dt ^\prime\nonumber\\ \times \langle0 | [j_\alpha ({\bf r}, t), \rho
({\bf r} ^ \prime, t ^\prime)] |0\rangle {\partial E_\beta ({\bf r} ^ \prime, t
^\prime) \over
\partial r_\gamma ^\prime},
\end{eqnarray}
and, similarly,
\begin{eqnarray}
\label {127} \langle 0 |\rho ({\bf r},t)|0\rangle^{(+)}&=
 &{i\over 2\hbar c} \int d {\bf r} ^ \prime r_\beta ^\prime r_\gamma ^\prime
\int _ {-\infty} ^tdt ^\prime
\nonumber\\
 &\times&\langle 0 | [\rho ({\bf r}, t), j_\delta ({\bf r} ^ \prime, t ^\prime)]
  |0\rangle{\partial^2a_\beta({\bf r} ^ \prime, t ^\prime) \over
\partial r_\gamma ^\prime\partial r_\delta ^\prime}\nonumber\\
&-& {i\over 2\hbar} \int d {\bf r} ^ \prime r_\beta ^\prime r_\gamma ^\prime
\int _ {-\infty} ^t dt ^\prime\nonumber\\
 &\times&\langle0|[\rho({\bf r}, t), \rho ({\bf r} ^ \prime, t ^\prime)] |0\rangle {\partial
E_\beta ({\bf r} ^ \prime, t ^\prime) \over \partial r_\gamma ^\prime}.
\end{eqnarray}
Let us notice, that unlike Eqs. (68) and (69), Eqs. (127) and (128) contain only
second derivative from a vector $ {\bf a} ({\bf r}, t) $. The obtained results can be
written down in the symmetric form:
\begin{eqnarray}
\label {128} \langle 0|j _ {1\alpha} ({\bf r}, t) |0\rangle ^
{(+)} = - {e\over 4mc} \langle
0 |\rho ({\bf r}) |0\rangle r_\beta r_\gamma\nonumber\\
\times {\partial\over \partial r_\alpha} (\partial a_\beta ({\bf r}, t) /\partial
r_\gamma + \partial a_\gamma ({\bf r}, t) /\partial r_\beta) + {i\over 4\hbar c} \nonumber\\
\times\int d {\bf r} ^ \prime r_\beta ^\prime r_\gamma ^\prime \int _ {-\infty} ^tdt
^\prime \langle 0 | [j_\alpha ({\bf r}, t), \Omega _ {\beta\gamma} ({\bf r} ^ \prime,
t ^\prime)] |0\rangle,
\end{eqnarray}
\begin{eqnarray}
\label {129} \langle 0 |\rho ({\bf r}, t) |0\rangle ^ {(+)} = {i\over 4\hbar c} \nonumber\\
\times \int d {\bf r} ^ \prime r_\beta ^\prime r_\gamma ^\prime \int _ {-\infty} ^t
dt ^\prime \langle 0 | [\rho ({\bf r}, t), \Omega _ {\beta\gamma} ({\bf r} ^ \prime,
t ^\prime)] |0\rangle
\end{eqnarray}
where the symmetric tensor
\begin{eqnarray}
\label {130} \Omega _ {\beta\gamma} ({\bf r},t)&=&j_\delta ({\bf
r}, t) {\partial\over
\partial r_\delta} (\partial a_\beta ({\bf r}, t) /\partial r_\gamma + \partial a_\gamma
( {\bf r}, t) /\partial r_\beta) \nonumber\\
 &+&\rho({\bf r}, t) {\partial\over \partial t} (\partial a_\beta ({\bf r}, t) /\partial
r_\gamma + \partial a_\gamma ({\bf r}, t) /\partial r_\beta).
\end{eqnarray}

\section{Constant magnetic field}

Let us consider a case of a constant in time and space magnetic fields $ {\bf H}
=const, {\bf E} =0 $. Let us remind, that we have included constant magnetic field in
unperturbed Hamiltonian $ {\cal H} $ (see section II). But in the present section we
do not include a field in our Hamiltonian and consider it so weak, that it is
possible to be limited by linear on field contributions to induced current and charge
densities. As $ {\bf E} =0 $, contributions to the average induced by $E $ densities
are equal 0 ( See Eqs. (66) and (67)). The contributions with indexes $
\partial E/\partial r $ are broken on two parts, which we have designated by the top
indexes (+) and (-). Parts with indexes (+) are equal 0. It is easy to be convinced of
it if to choose vector and scalar potentials, for example, in gage \cite{5}
\begin{equation}
\label {131} { \bf A} ({\bf r}) = (1/2) [{\bf r} \times {\bf H}],
\quad \varphi =0
\end{equation}
and to use Eqs. (118), (126) and (127).

So, there are only contributions with indexes (-), determined in Eqs. (122) and (123).
Having put $ {\bf H} ({\bf r}, t) = {\bf H} $ and having made replacement variable $t
^\prime $ by $t ^ {\prime\prime} =t ^\prime -t $, we obtain time-independent results:
\begin{eqnarray}
\label {132} \langle0|j _ {1\alpha} ({\bf r}, t) |0\rangle_H = -
{e\over 2mc} ({\bf
H} \times {\bf r}) _ \alpha \langle0 |\rho ({\bf r}) |0\rangle + {i\over 2\hbar c} \nonumber\\
\times \int d {\bf r} ^ \prime \int _ {-\infty} ^0dt ^ {\prime\prime} ({\bf H} \times
{\bf r} ^ \prime) _ \beta \langle 0 | [j_\alpha ({\bf r}), j_\beta ({\bf r} ^ \prime,
t ^ {\prime\prime})] |0\rangle,
\end{eqnarray}
\begin{eqnarray}
\label {133} \langle0 |\rho_1 ({\bf r}, t) |0\rangle_H = {i\over 2\hbar c} \nonumber\\
\times \int d {\bf r} ^ \prime \int _ {-\infty} ^0dt ^ {\prime\prime} ({\bf H} \times
{\bf r} ^ \prime) _ \beta \langle 0 | [\rho ({\bf r}), j_\beta ({\bf r} ^ \prime, t ^
{\prime\prime})] |0\rangle.
\end{eqnarray}
The indexes $H $ in the LHSs mean $ {\bf H} =const, \, {\bf E} =0 $.

In obtained expressions we shall pass from operators ${\bf r}_i$
to the operators ${\bar {\bf r}}_i$, determined in Eq. (81). By a
way, similar to a way stated in section X, we obtain
\begin{eqnarray}
\label{134} \langle0|j_{1\alpha}({\bf r},t)|0\rangle_H= -{e\over
2mc}({\bf
H}\times\langle 0|\bar{\bf d}({\bf r})|0\rangle)_\alpha\nonumber\\
+{ie\over 2\hbar c}\int_{-\infty}^0dt^\prime \sum_i\{\langle
0|[j_\alpha({\bf r},-t^\prime),({\bf H}\times{\bar{\bf
r}}_i)_\beta
 v_{i\beta}]|0\rangle,
\end{eqnarray}
\begin{eqnarray}
\label{135} \langle0|\rho_1({\bf r},t)|0\rangle_H={ie\over 2\hbar c}\nonumber\\
\times \int_{-\infty}^0dt^\prime\sum_i\{\langle 0| [\rho ({\bf
r},-t^\prime), ({\bf H}\times{\bar{\bf r}}_i)_\beta
 v_{i\beta}]|0\rangle.
\end{eqnarray}

Let us note that the relations are performed
$$div \langle0|{\bf j}_1({\bf r},t)|0\rangle_H=0,\, \int \langle0|{\bf
j}_1({\bf r},t)|0\rangle_H\,d{\bf r}=0.$$

\section{Conclusion}

Let us consider the basic results. Since operator of interaction of charged particles
with electromagnetic field is expressed through potentials $ {\bf A} ({\bf r}, t) $
and $ \varphi ( {\bf r}, t) $, but not through electric $ {\bf E} ({\bf r}, t) $ and
magnetic $ {\bf H} ({\bf r}, t) $ fields, initial expressions for average values of
induced  current and charge densities also are expressed through potentials (see Eqs.
(17) and (18)). But Eq. (17) for  a current density is inconvenient, since it
contains the contribution proportional to average $ \langle0 |\rho ({\bf r}, t)
|0\rangle $ of a charge density, and this contribution is not completely small.

Therefore we have put a task to express induced densities through fields $ {\bf E}
({\bf r}, t) $ and $ {\bf H} ({\bf r}, t) $, that is certainly feasible, since the
average current and charge densities are observable values.

In result Eqs. (31) and (32) were obtained, containing only electric field. However,
Eq. (31) has the same lack, as Eq. (17), - it contains the contribution proportional $
\langle0 |\rho ({\bf r}, t) |0\rangle $.

If to use an approximation, in which the electric field $ {\bf E} ({\bf r}, t) $ does
not depend on coordinates, but only on time, the contribution proportional $ \langle 0
|\rho ({\bf r}, t) |0\rangle $, may be removed, if to enter in expression for
average  current density the operators $ {\bf r}_i $ of $ i $th particle coordinates
, having taken  a ratio $ {\bf v}_i = (i/\hbar)[{\cal H}, {\bf r}_i] $. This
reception in essence is used in \cite{3}, where the theory of light scattering in bulk
crystals is discussed. In the appendix A it is shown, how the same reception allows
to exclude  the charged particles concentration  from the formula for conductivity of
intrinsic bulk semiconductors at $T=0$.

The same result turns out, if in approximation $ {\bf E} ({\bf r}, t) \simeq {\bf E}
(t) $ to write down interaction of particles with a field as $ -e\sum_ir _ {i\beta}
E_\beta (t) $, as it is made in \cite{8}.

However, the task becomes complicated if to take into account dependence of an
electric field on coordinates. We have put to ourselves by the purpose to obtain such
expression for induced  current density, which would pass in the Kubo-formula at
transition from $ {\bf E} ({\bf r}, t) $ to $ {\bf E} (t) $, and similar expressions
for induced  charge density. For a case $T=0 $ this task is solved in sections VII -
IX.

Essence of used reception consists in the following. The operator of interaction of
particles with a field, expressed through electric and magnetic fields, does not
exist. But we enter a fictitious operator of interaction, which results into correct
results for average on the ground state of system  values of the induced current and
charge densities. This fictitious operator $U_1^f $ Eq. (37) is expressed only through
an electric field $ {\bf E} ({\bf r}, t) $, in an integral on time. Then such
transformation of interactions $U_1^f\to {\tilde U} _1 $ is made, which does not
change values  of average densities, but eliminate the contribution containing  $
\langle0 |\rho ({\bf r}, t) |0\rangle $ from expression for average induced current
density. In result  Eqs (64) - (69) for $ \langle0 | {\bf j} _1 ({\bf r}, t)
|0\rangle $ and $ \langle0 |\rho_1 ({\bf r}, t) |0\rangle $ are obtained, in which the
basic contributions contain an electric field, and additional - the derivatives from
$ {\bf E} ({\bf r}, t) $ on coordinates. The transformed  fictitious interaction $
{\tilde U} _1 $ also is divided on basic and additional parts. The first is equal $
-e\sum_ir _ {i\beta} E_\beta ({\bf r} _i, t) $, the second contains derivative $
\partial E_\alpha ({\bf r}, t) /
\partial {\bf r} _ \beta $ in an integral on time.

In section X Eqs. (64) - (69) for average densities of induced
currents and charges are transformed so, that obviously do not
depend on an origin of coordinates  ${\bf r}_i$ of particles. It
is the basic result of the present work. In section X from Eqs.
(64) - (69) for average  induced currents and charges densities
the diagonal elements  $ \langle 0 | {\bf r} _i|0\rangle $ are
excluded. The results Eqs. (90) - (96), containing only
non-diagonal elements of the operators $ {\bf r} _i $, are
obtained . This is the main result of the present work. In
section XI the expressions for the conductivity tensor $ \sigma
({\bf k}, \omega | {\bf r}) $ are obtained, dependent from
coordinates, in a case of spatially-heterogeneous systems.

Further in section XII from expressions for $ \langle0 | {\bf j} _1 ({\bf r}, t)
|0\rangle $ and $ \langle0 |\rho_1 ({\bf r}, t) |0\rangle $, containing only electric
field and derivative from i) on coordinates, we pass to expressions containing also
magnetic field $ {\bf H} ({\bf r}, t) $. It is necessary for reception of results in
case $ {\bf H} =const, {\bf E} =0 $, which is considered in section XIII.

 In the Appendix B the expression
for the operator of total acceleration of system of the charged
particles is obtained. The acceleration is caused by an external
weak electromagnetic field. It is shown, that averaged on to the
basic state of system the total acceleration can be expressed
through  electrical and magnetic fields. In the case of free
particles the obtained result passes to a correct limit containing
a force caused by an electric field, and the Lorentz force.
Further it is supposed to use obtained results for construction
of the general theory of secondary radiation from low-dimensional
semiconductor objects.

 The work has obtained the partial financial support from
 the Russian Fund of basic researches
 (00-02-16904), Program MNTK " Physics of semiconductor nano-structures"
and the Federal Program "Integration".

\begin{center}
{ \bf APPENDIX A}

{\bf Three expressions for the conductivity tensor  in a case of a spatially -
homogeneous matter and an electric field independent on coordinates.}
\end{center}

Proceeding from Eq. (31) for the average  induced current density and taking into
account a following connection between the Fourier-components
$$ ~~~~~~~~~~~~~~~ {\bf a} ({\bf k}, \omega) = (ic/\omega) {\b E} ({\bf k}, \omega),
~~~~~~~~~~~~~~~~~~~ (A1) $$ following from definition Eq. (30), we obtain
$$\sigma _ {\alpha\beta} ({\bf k}, \omega | {\bf r}) =
{ ie\over m\omega} \langle 0 |\rho ({\bf r}) |0\rangle\delta _ {\alpha\beta} +
{1\over\hbar\omega} \int d {\bf r} ^ \prime\int _ {-\infty} ^ \infty dt ^\prime
$$
$$\times e ^ {-i {\bf k} {\bf r} ^ \prime +i\omega t ^\prime} \Theta (t ^\prime)
\langle 0|j_\alpha ({\bf r}), j_\beta ({\bf r} - {\bf r} ^ \prime, -t ^\prime)
|0\rangle. ~~~~~~~~ (A2) $$ {\it In a case of spatially - homogeneous system} the
tensor $ \sigma _ {\alpha\beta} ({\bf k}, \omega | {\bf r}) $ does not depend on $
{\bf r}. $ In the first term we use a ratio
$$ ~~~~~~~~~~~~~~~~~~~~~~~~~~~\langle 0 |\rho ({\bf r}) |0\rangle=en,
~~~~~~~~~~~~~~~~~~~~~~~~~~~~~ (A3) $$ where $n $ is the concentration of charged
particles, in the second term we integrate on $ {\bf r} $ and $ {\bf r} ^ \prime $,
also we shall divide the result on the normalized volume $V_0 $. We obtain
$$\sigma _ {\alpha\beta} ({\bf k}, \omega) = {ie^2n\over m\omega} \delta _ {\alpha\beta}
+ {e^2\over 4m^2\hbar\omega V_0} \sum _ {i, j} \int _ {-\infty} ^ \infty dt \Theta (t)
e ^ {i\omega t} $$
 $$\times\left\langle0\left|\left[\left\{e^{i{\cal H} t/\hbar} (e ^ {-i {\bf k} {\bf r} _i}
 p_{i\alpha}\right.\right.\right.\right.$$
 $$\left.\left.\left.\left.+p_{i\alpha}e^{-i{\bf k} {\bf r} _i})
e ^ {-i {\cal H} t/\hbar} \right \}, (e ^ {i {\bf k} {\bf r} _j} p _ {j\beta} + p _
{i\beta} e ^ {i {\bf k} {\bf r} _j}) \right] \right|0\right\rangle, ~~~~~~ (A4), $$
where $ {\bf p} _i $ is the momentum operator determined in Eq. (4) with the account
of a constant strong magnetic field.

{ \it In case of electric field} $ {\bf E} (t) $, {\it independent on coordinates},
we enter the frequency representation
$$ ~~~~~~~~~~ E_\alpha (\omega) =
\int _ {-\infty} ^ \infty dt E_\alpha (t) \exp (i\omega t), ~~~~~~~~~~~~~~~ (A5) $$
and
$$ ~~~~~~~ E_\alpha ({\bf k}, \omega) =
(2\pi) ^3\delta ({\bf k})E_\alpha(\omega).~~~~~~~~~~~~~~~(A6).$$ Let us enter also
conductivity $ \sigma _ {\alpha\beta} (\omega | {\bf r}) $, included in definition of
an average current density
 $$\langle0|j_{1\alpha}({\bf r}, t)|0\rangle_h^+={1\over2\pi}\int_0^\infty d\omega
\sigma _ {\alpha\beta} (\omega | {\bf r}) E_\alpha (\omega) e ^ {-i\omega t}, ~~~
(A7) $$ where the index $h $ means spatially-homogeneous electric field.

Using section XI, it is easy to show, that
$$\sigma _ {\alpha\beta} (\omega | {\bf r}) =
\sigma _ {\alpha\beta} ({\bf k} =0, \omega | {\bf r}). ~~~~~~~~~ (A8) $$ From (A.4)
and (A.8) in a case of spatially - homogeneous matter and the field $ {\bf E} (t) $ we
obtain
$$\sigma _ {\alpha\beta} ^I (\omega) = {ie^2n\over m\omega}
\delta _ {\alpha\beta} $$
$$ + {e^2\over \hbar\omega V_0} \int _ {-\infty} ^ \infty dt
\Theta (t) e ^ {i\omega t}\langle0|[V_\alpha(t),V_\beta]|\rangle,~~~~(A9)$$ where
$V_\alpha = (1/m) \sum_ip _ {i\alpha} $ is the operator of total speed of charged
particles. (A9) is the  first formula for the conductivity tensor. For reception two
others we use a ratio
$$ ~~~~~~ V_\alpha = (i/\hbar) [{\cal H}, R_\alpha], ~~~~~~~~~~~~~~~ (A10) $$
where $R_\alpha =\sum_ir _ {i\alpha}. $ Having substituted (A10) in (A8), we obtain
$$\sigma _ {\alpha\beta} (\omega) = {ie^2n\over m\omega}
\delta _ {\alpha\beta} + {e^2\over \hbar\omega V_0} \int _ {-\infty} ^ \infty dt
\Theta (t) $$
$$\times e ^ {i\omega t} {d\over dt}\langle0|[R_\alpha(t),V_\beta]|0\rangle.
~~~~~~~~~~~~~~~~~~ (A11) $$ Let us execute integration on $t $ in parts. At
$t\to\infty $ function $ \Theta (t) e ^ {i\omega t} {d\over dt}
 \langle0|[R_\alpha(t),V_\beta]|0\rangle\to 0 $ at replacement $ \omega $ by
$ \omega +i\delta $, $ \delta\to 0 $. Let us take into account also, that $d\Theta (t)
/dt =\delta (t) $ and
 $$~~~~~~~~~(1/V_0)[R_\alpha,V_\beta]=(i\hbar n/m) \delta _ {\alpha\beta},
~~~~~~~~~~~ (A12) $$ then we shall obtain
 $$\sigma_{\alpha\beta}^{II}(\omega)=-{ie\over \hbar V_0}
\int _ {-\infty} ^ \infty dt\Theta (t) e ^ {i\omega t} \langle 0
 |[D_\alpha(t),V_\beta]|0\rangle,~~(A13)$$
where the designation is entered
$$ ~~~~~~~~~~~~~~~~~~~~~~~ {\bf D} =e {\bf R}. ~~~~~~~~~~~~~~~~~~~ (A14). $$
Eq. (A13) is the second expression for the conductivity tensor. Let us emphasize, that
at transition from (A9) to (A13) there was a reduction of the first term from the RHS
(A9), containing concentration $n $.

At transition to the third expression  (A10) for the operator $V_\beta $ from the RHS
(A13) is used. Further we work under the same circuit, as at transition from (A9) to
(A13). Taking into account, that the commutator $ [D_\alpha, D_\beta] =0 $, we obtain
the third formula
 $$\sigma_{\alpha\beta}^{III}(\omega)={\omega\over \hbar V_0} \int _ {-\infty} ^ \infty dt
\Theta (t) e ^ {i\omega t} \langle0 | [D_\alpha (t), D_\beta] |0\rangle. ~~~ (A15). $$

It is convenient to apply  (A9), (A13) and (A15) to different
systems.

For free particles, when $V ({\bf r} _1... {\bf r} _N) =0, \; {\bf H} _c=0 $, the
velocity operator $ {\bf v} $ commutes with the Hamiltonian $ {\cal H} _ {free}
=m\sum_iv_i^2. $ Let us use (A 9), taking into account, that $v_\alpha (t) =v_\alpha,
\; [v_\alpha, v_\beta] =0, $ and we obtain the known result
$$\sigma _ {\alpha\beta \, free} (\omega) =
{ ie^2n\over m\omega}\delta_{\alpha\beta}.~~~~~~~~~~~~~(A.16)$$

In a case, when exited states are separated from the ground state by an energy gap ,
i. e. the energy of these states $E_n =\hbar\omega_n\neq 0 $, it is more convenient
to use (A 13) or (A 15). Using exact wave functions $ |n > $ of the exited states and
having calculated integral on time, we shall express (A 15) through matrix elements of
the operator $ {\bf D} $:
 $$\sigma_{\alpha\beta}^{III}(\omega)={i\omega\over\hbar V_0}
\sum_n\left \{{< 0|D_\alpha |n > <n|D_\beta|0> \over \omega - \omega_n} -\right. $$
$$\left. {< 0|D_\beta |n > < n|D_\alpha|0 > \over \omega +
\omega_n} \right \}, ~~~~~~~ (A.17) $$ whence follows, that at $ \omega\to 0 \;\sigma
_ {\alpha\beta} (\omega) \to 0 $ for systems with an energy gap (superconductivity is
not considered). To number of systems with an energy gap the bulk semiconductors
without impurity and defects concern.

The conductivity tensor $ \sigma _ {\alpha\beta} (\omega) $ is connected with the
dielectric susceptibility tensor $ \varepsilon _ {\alpha\beta} (\omega) $ by ratio
 $$\varepsilon_{\alpha\beta}(\omega)=\delta_{\alpha\beta}+
4\pi\chi _ {\alpha\beta} (\omega), $$
 $$\chi_{\alpha\beta}(\omega)=(i/\omega)\sigma_{\alpha\beta}(\omega),$$
so from (A.17) we obtain
 $$\chi_{\alpha\beta}(\omega)={1\over\hbar V_0}
\sum_n\left \{{< 0|D_\alpha |n > <n|D_\beta|0> \over \omega_n - \omega}\right. $$
$$+\left. {< 0|D_\beta |n > < n|D_\alpha|0 > \over \omega_n +
\omega} \right \}. ~~~~~~~ (A.18) $$ Using the ratio between matrix elements
 $$<0|D_\alpha|n>=(i\,e/\omega_n)<0|v_\alpha|n>,$$
 $$<n|D_\alpha|0>=-(i\,e/\omega_n)<n|v_\alpha|0>,$$
from (And 18) we find
 $$\chi_{\alpha\beta}(\omega)={e^2\over\hbar V_0}
 \sum_n{1\over\omega_n^2}\left\{{<0|v_\alpha|n><n|v_\beta|0>\over \omega_n -
\omega}\right. $$
$$+\left. {< 0|v_\beta |n > < n|v_\alpha|0 > \over \omega_n +
\omega} \right \}. ~~~~~~~ (A.19) $$

If $ \omega\ll\omega_n $, $ \chi _ {\alpha\beta} $ is real and also does not depend
on frequency $ \omega $, but at $ \omega\simeq\omega_n $ this dependence becomes
strong and there appears an imaginary part, distinct from zero, of the tensor $ \chi _
{\alpha\beta} $, determining the resonant light absorption on frequencies $
\omega\simeq\omega_n $. For calculation of an imaginary part it is necessary to
replace frequency $ \omega $ by $ \omega+i\gamma/2, \; \gamma\to 0 $, when the field
is switched on adiabatically, or to replace frequency $ \omega_n $ by $
\omega_n-i\gamma/2 $, i. e. to take into account a final life time of a system in a
state $n $.

\begin{center}
{\bf APPENDIX B}

{\bf Acceleration of particles system }
\end{center}

To obtain of the operator $ {\bf J} _1 (t) $ of the induced current of system of
particles we use Eq. (14), containing vector and scalar potentials. Having executed
integration on $ {\bf r} $ and $ {\bf r} ^ \prime $, we obtain
$$ J _ {1\alpha} (t) = - {e^2\over mc} \sum_iA_\alpha ({\bf r} _i (t), t) + $$
 $$~~~~~+{ie\over\hbar}\int_{-\infty}^\infty dt ^\prime
 \left[U_1(t^\prime),\sum_iv_{i\alpha}(t)\right],~~~~~~~~~~~~(B1)$$ where
the designation is used
$$ {\bf A} ({\bf r} _i (t), t) =
e ^ {i {\cal H} t/\hbar} {\bf A} ({\bf r} _i, t) e ^ {-i {\cal H} t/\hbar}. $$
Differentiating (B1) on time and having divided on $e $, we shall obtain the operator
$ {\bf W} _1 (t) $ of the induced total acceleration:
$$ W _ {1\alpha} (t) = - {e\over mc} {d\over dt} \sum_iA_\alpha ({\bf r} _i (t), t) +
{ i\over \hbar}\left[U_1(t),\sum_iv_{i\alpha}(t)\right]$$
$$ ~~~~~~ + {i\over \hbar} \int _ {-\infty} ^t dt ^\prime
[ U_1(t^\prime),W_\alpha(t)],~~~~~~~~~(B2)$$ where
$W_\alpha(t)=\sum_iw_{i\alpha}(t),~~w_{i\alpha}(t)=dv_{i\alpha}(t)/dt.$ The decoding
of first two terms gives
$$ {\bf W} _ {1\alpha} (t) = - {e\over mc} \sum_i {\partial A\alpha ({\bf r} _i (t), t)
\over \partial t} - - {e\over m} \sum_i {\partial \varphi ({\bf r} _i (t), t) \over
\partial {\bf r} _ {i\alpha} (t)} $$
$$ + {e\over 2mc} \sum_i\left \{v _ {i\alpha} (t) \left
( {\partial A_\beta ({\bf r} _i (t), t) \over \partial {\bf r} _ {i\alpha} (t)} -
{\partial A_\alpha ({\bf r} _i (t), t) \over \partial {\bf r} _ {i\beta} (t)} \right)
\right.
$$
$$ + \left.\left ({\partial A_\beta ({\bf r} _i (t), t) \over \partial {\bf r} _ {i\alpha} (t)} -
{ \partial A_\alpha ({\bf r} _i (t), t) \over \partial {\bf r} _ {i\beta} (t)} \right)
v _ {i\alpha} (t) \right \} $$
$$ ~~~~~- (e/mc) ^2\sum_i [{\bf A} ({\bf r} _i (t), t) \times {\bf H} _c] _ \alpha $$
$$ ~~~~ + (i/\hbar) \int _ {-\infty} ^t dt ^\prime [U_1 (t ^\prime), W_\alpha (t)],
~~~~~~~~~~~ (B3) $$ where the designations are used:
$$ {\partial A\alpha ({\bf r} _i (t), t) \over \partial t} =
e ^ {i {\cal H} t/\hbar} {\partial A_\alpha ({\bf r} _i, t) \over \partial t} e ^ {-i
{\cal H} t/\hbar}, $$
$$ {\partial \varphi ({\bf r} _i (t), t) \over \partial {\bf r} _ {i\alpha} (t)} =
e ^ {i {\cal H} t/\hbar} {\partial \varphi ({\bf r} _i, t) \over
\partial {\bf r} _ {i\alpha}} e ^ {-i {\cal H} t/\hbar}, $$
$ {\bf H} _c $ is the constant strong magnetic field included in the basic Hamiltonian
$ {\cal H} $. At transition from (B2) to (B3) the ratio is used $ {\bf v} _i\times
{\bf v} _i = (i\hbar e/m^2c) {\bf H} _c, $ and the LHS $ \neq 0 $ because of
non-commutativity of various projections of the velocity, for example
$$ [{\bf v} _i\times {\bf v} _i] _z = [v _ {ix}, v _ {iy}] = (i\hbar e/m^2c) H _ {cz}.
~~~~~~~~~~ (B4) $$ The operator $ {\bf w} _i (t) $ of acceleration $i$th particle is
equal
$$ w _ {i\alpha} = {i\over \hbar} e ^ {i {\cal H} t/\hbar} [{\cal H}, v _ {i\alpha}]
e ^ {-i {\cal H} t/\hbar} = $$
$$ = - {1\over m} e ^ {i {\cal H} t/\hbar}
{ \partial V ({\bf r} _1, \dots {\bf r} _N) \over
\partial r _ {i\alpha}} e ^ {-i {\cal H} t/\hbar} $$
$$ + {e ({\bf v} _i (t) \times {\bf H} _c) _ \alpha\over mc}.
~~~~~~~~~~~~~~~~~~ (B5) $$ With the help Eq. (1) it is easy to see, that the
expression in braces in (B3) is equal
$$ e ^ {i {\cal H} t/\hbar} \{({\bf v} _i\times {\bf H} ({\bf r} _i, t)) _ \alpha-
( {\bf H} ({\bf r} _i, t) \times {\bf v} _i) _ \alpha \} e ^ {-i {\cal H} t/\hbar}. $$
Therefore (B3) will be transformed to
$$ {\bf W} _1 (t) = {e\over m} \sum_i {\bf E} ({\bf r} _i (t), t) $$
$$ + {e\over 2mc} \sum_i \{{\bf v} _i (t) \times {\bf H} ({\bf r} _i (t), t) -
{ \bf H} _i ({\bf r} _i (t), t) \times {\bf v} _i (t) \} $$
$$-\left ({e\over mc} \right) ^2\sum_i {\bf A} ({\bf r} (t), t) \times {\bf H} _c $$
$$ + {i\over \hbar} \int _ {-\infty} ^t dt ^\prime [U_1 (t ^\prime), {\bf W} (t)].
~~~~~~~~~~~~~~~~~~~ (B6) $$ It is obvious, that the second term corresponds to the
Lorentz force, at which the non-commutativity некоммутативность of the operators $
{\bf v} _i $ and $ {\bf H} ({\bf r} _i, t) $ is taken into account. The third term
caused by $ {\bf H} _c $, contains the additive to speed $ \Delta {\bf v} _i $,
determined in Eq. (8) and induced by a weak electromagnetic field.

In the case of free particles
$$ {\bf H} _c=0, \quad V ({\bf r} _1, \dots {\bf r} _N) =0, \quad {\bf W} =0, ~~~~~~~~~ (B7) $$
and in the RHS (B6) only two first terms containing weak electric and magnetic fields
are kept .

But if particles are not free, the operator (B6) fails to be expressed only through
fields, since the last two terms  contain vector and scalar potentials. Average value
$ \langle 0 | {\bf W} _1 (t) |0\rangle $ of the induced acceleration should be
expressed only through fields, that we now shall prove. For this purpose we shall
calculate the value $ \langle 0 | {\bf W} _1 (t) |0\rangle $ in another way, and then
we shall check up, that both ways give coinciding results. Let us use Eq. (27) and
integrate its both parts on $ {\bf r} $. We obtain
$$\langle 0|W _ {1\alpha} (t) |0\rangle = {e\over m} \sum_i
\langle 0 | E_\alpha ({\bf r} _i, t) |0\rangle $$
$$- {ie\over 2\hbar} \int _ {-\infty} ^0dt ^\prime
\left\langle 0\left|\left[\sum_jv_{j\beta},\sum_i\{v_{i\beta}(t^\prime E_\beta ({\bf
r}_i(t^\prime),t+t^\prime)\right.\right.\right.$$
$$ +E_\beta ({\bf r}_i(t^\prime),t+t^\prime)v_{j\beta}(t^\prime)\}\bigg]
\bigg|0\bigg\rangle. ~~~~~~~~~~~~~~~~~ (B8) $$ This expression contains only electric
fields. On the other hand, having averaged the operator (B6), we obtain
$$\langle 0|W _ {1\alpha} (t) |0\rangle = {e\over m} \sum_i
\langle 0 | E_\alpha ({\bf r} _i, t) |0\rangle $$
$$ + {e\over 2mc} \sum_i\langle 0 | ({\bf v} _i\times {\bf H} ({\bf r} _i, t)) _ \alpha-
( {\bf H} ({\bf r} _i, t) \times {\bf v} _i) _ \alpha|0\rangle $$
$$- (e/mc) ^2\sum_i\langle 0 | ({\bf A} ({\bf r} _i, t) \times {\bf H} _c) _ \alpha |
 0\rangle+C_\alpha(t),~~~~~~~~~~~~~~~~(B9)$$
where
$$ C_\alpha (t) = (i/\hbar) \int _ {-\infty} ^t dt ^\prime\langle0 |
[ U_1(t^\prime),W_\alpha(t)]|0\rangle.~~~~~~~~~~~~~~(B10)$$ Let us transform (B10).
Integrating in parts, we obtain
 $$~~~~C_\alpha(t)=C_\alpha^1(t)+C_\alpha^2(t),~~~~~~~~~~~~~~~~~~~~~(B11)$$
 $$C_\alpha^1(t)=-(i/\hbar)\left\langle0\left|\left[U_1,\sum_iv_{i\alpha}\right]\right|0
\right\rangle, ~~~~~~~~~~~~ (B12) $$
 $$ ~~~~ C_\alpha^2 (t) = {i\over \hbar} {d\over dt} \int _ {-\infty} ^t dt ^\prime
 \left\langle0\left|\left[U_1(t^\prime),\sum_iv_{i\alpha}(t)\right]\right|0\right\rangle.
~~~~ (B13) $$ In the RHS (B12) we shall substitute Eq. (5) for $U_1 $ and we shall
calculate the commutator. We obtain
$$ C_\alpha^1 (t) = {e\over m} \sum_i\left\langle0\left | {\partial \varphi ({\bf r} _i, t)
\over \partial r _ {i\alpha}} \right|0\right\rangle $$
$$ + \left ({e\over mc} \right) ^2
\sum_i\langle0 | ({\bf A} ({\bf r} _i, t) \times {\bf H} _c) _ \alpha|0\rangle $$
$$- {e\over 2mc} \sum_i\left\langle0\left|v _ {i\beta}
{ \partial A_\beta ({\bf r} _i, t) \over\partial r _ {i\alpha}} + {\partial A_\beta
({\bf r} _i, t) \over\partial r _ {i\alpha}} v _ {i\beta} \right|0\right\rangle. ~~
(B14) $$

In (B13) in integral we pass to $t ^ {\prime\prime} =t-t ^\prime $ and we break the
integral on two parts, first of which contains vector, and second - scalar potentials:
 $$C_\alpha^2(t)=C_{A\alpha}^2(t)+C_{\varphi\alpha}^2(t),~~~~~~~~~~~~~~~~~(B15)$$
$$ C _ {A\alpha} ^2 (t) = {ie\over 2c\hbar} \int _ {-\infty} ^0dt ^\prime
 \left\langle0\left|\left[\sum_jv_{j\alpha},\right.\right.\right.$$
 $$\times\sum_i\left\{v_{i\beta}(t^\prime)
{ \partial A_\beta ({\bf r} _i (t ^\prime), t+t ^\prime) \over\partial t} \right. $$
$$\left.\left.\left.\left. + {\partial A_\beta ({\bf r} _i (t ^\prime), t+t ^\prime) \over\partial t}
 v_{i\beta}(t^\prime)\right\}\right]\right|0\right\rangle,~~~~~~~~~~~~~~~(B16)$$
$$ C _ {\varphi\alpha} ^2 (t) = - {ie\over \hbar} \int _ {-\infty} ^0dt ^\prime $$
$$\times\left\langle0\left |\left [\sum_jv _ {j\alpha}, \sum_i
{ \partial \varphi ({\bf r} _i (t ^\prime), t+t ^\prime) \over \partial t} \right ]
\right|0\right\rangle.~~~~~~~~~~~~~~~~(B17)$$ We shall leave (B16) without changes,
and in (B17) we use a ratio
$$\partial\varphi ({\bf r} _i (t ^\prime), t+t ^\prime) /\partial t = $$
$$ = \exp (i {\cal H} t ^\prime/\hbar)
( \partial\varphi ({\bf r} _i, t+t ^\prime) /\partial t) \exp (-i {\cal H} t
^\prime/\hbar) = $$
$$ = \exp (i {\cal H} t ^\prime/\hbar)
( \partial\varphi ({\bf r} _i, t+t ^\prime) /\partial t ^\prime) \exp (-i {\cal H} t
^\prime/\hbar). $$ Then we integrate in parts on $t ^\prime $ and  obtain
$$ C _ {\varphi\alpha} ^2 (t) = - {ie\over \hbar} \sum_i\langle0 | [v _ {i\alpha},
\varphi ({\bf r} _i, t)] |0\rangle $$
$$- {e\over \hbar^2}\int_{-\infty}^0dt^\prime\left\langle0\left|\left[\sum_j
v _ {j\alpha}, e ^ {i {\cal H} t ^\prime/\hbar} \right.\right.\right. $$
$$\left.\left.\left.\times\sum_i [{\cal H},
\varphi ({\bf r} _i, t+t ^\prime)] e ^ {-i {\cal H} t ^\prime/\hbar} \right]
\right|0\right\rangle. ~~~~~~~ (B18) $$ The commutator in the first term is
calculated, and for transformations of the second term we notice, that
$$ {i\over \hbar} [{\cal H}, \varphi ({\bf r} _i, t)] = {1\over 2}
\left (v _ {i\beta} {\partial \varphi ({\bf r} _i, t) \over \partial r _ {i\beta}} +
{ \partial \varphi ({\bf r} _i, t) \over \partial r _ {i\beta}} v _ {i\beta} \right).
~ (B19) $$ Having substituted (B19) in (B18), we obtain:
$$ C _ {\varphi\alpha} ^2 (t) = - {e\over m} \sum_i\langle0 |
\partial \varphi ({\bf r} _i, t) /\partial r _ {i\alpha} |0\rangle + $$
$$ + {ie\over 2\hbar} \int _ {-\infty} ^0dt ^\prime
\left\langle0\left |\left [\sum_j v _ {j\alpha},
 \sum_i(v_{i\beta}(t^\prime){ \partial \varphi ({\bf r} _i (t ^\prime), t+t ^\prime) \over
\partial r _ {i\beta} (t ^\prime)}\right.\right.\right.$$
$$\left.\left.\left.
 + {\partial \varphi
( {\bf r} _i (t ^\prime), t+t ^\prime) \over\partial r _ {i\beta} (t ^\prime)}
 v_{i\beta}(t^\prime))\right]\right|0\right\rangle.~~~~~(B20)$$ Summing (B14), (B16)
and (B20), we obtain finally
$$ C_\alpha (t) = - {e\over 2mc}\left\langle0\left|v_{i\beta}{\partial A_\beta ({\bf r} _i, t)
\over\partial r _ {i\alpha}} + {\partial A_\beta ({\bf r} _i, t) \over
\partial r _ {i\alpha}} v _ {i\beta} \right|0\right\rangle $$
$$ + (e/mc) ^2\sum_i\langle0 | ({\bf A} ({\bf r} _i, t) \times {\bf H} _c) _ \alpha
|0\rangle- {ie\over 2\hbar} \int _ {-\infty} ^0dt ^\prime $$
 $$\times\left\langle0\left|\left[\sum_jv_{j\alpha},\sum_i\{
v _ {i\beta} (t ^\prime) E_\beta ({\bf
r}_i(t^\prime),t+t^\prime)\right.\right.\right.$$
$$ +E_\beta ({\bf r}_i(t^\prime),t+t^\prime)v_{i\beta}(t^\prime)\}\Bigg]
\Bigg|0\Bigg\rangle. ~~~~~~~~~~~~~~~ (B21) $$ Having substituted (B21) in (B9), we
obtain
$$\langle0|W _ {1\alpha} |0\rangle = {e\over m} \sum_i
\langle0|E_\alpha ({\bf r} _i, t) |0\rangle- {ie\over
 2\hbar}\int_{-\infty}^0dt^\prime$$
 $$\left\langle0\left|\left
[ \sum_jv_{j\alpha}, \sum_i\{v_{i\beta}(t^\prime)E_\beta({\bf r}_i
 (t ^\prime), t+t ^\prime)\right.\right.\right.$$
 $$\left.\left.\left. +
E_\beta ({\bf
 r}_i(t^\prime),t+t^\prime)v_{i\beta}(t^\prime)\}\right]\right|0\right\rangle$$
$$- {e\over 2mc}\sum_i\left\langle0\left|v_{i\beta}{\partial A_\alpha ({\bf r} _i, t)
\over \partial r _ {i\beta}} + {\partial A_\alpha ({\bf r} _i, t) \over
\partial r _ {i\beta}} v _ {i\beta} \right|0\right\rangle. ~ (B22) $$
This expression coincides with (B8) except for last term  from the
RHS (B22). But it is possible to show, that this term is equal 0.
Really,
 $$(1/2)\left\langle0\left|v_{i\beta}{\partial A_\alpha ({\bf r} _i, t) \over
\partial r _ {i\beta}} + {\partial A_\alpha ({\bf r} _i, t) \over
\partial r _ {i\beta}} v _ {i\beta} \right|0\right\rangle = $$
$$ = (i/\hbar) \langle0 | [{\cal H}, A_\alpha ({\bf r} _i, t)] |0\rangle=0,
~~~~~~~~~~~~ (B23) $$ since the operator $ {\cal H} $ has only diagonal matrix
elements $ \langle0 | {\cal H} |0\rangle $. So, we have checked up, {\it that results
(B8) and (B9), obtained by different ways, coincide}.

Let us consider a case of free particles, when the conditions (B7) are carried out.
Then $C_\alpha (t) =0 $, the penultimate term from the RHS (B9) also is equal to zero,
and comparing (B8) and (B4), we find, that for the average Lorentz force  a ratio
owes to be carried out
$$ {e\over 2c} \langle0 | [{\bf v} _i\times {\bf H} ({\bf r} _i, t)]) _ \alpha
- [{\bf H} ({\bf r} _i, t) \times {\bf v} _i] _ \alpha|0\rangle _ {free} = $$
$$ = - {iem\over 2\hbar}\int_{-\infty}^0dt^\prime\left\langle0\left|\left[\sum_jv_{j\alpha},
\sum_i \{v _ {i\beta} (t ^\prime) E_\beta ({\bf
 r}_i(t^\prime),t+t^\prime)\right.\right.\right.$$
$$ +E_\beta ({\bf r}_i(t^\prime),t+t^\prime)v_{i\beta}(t^\prime)\}\bigg]
\bigg|0\bigg\rangle _ {free}, ~~~~~~~~~~~~~~~~ (B24) $$ where the index $ "free" $
means performance of conditions (B7). Let us check up (B24) by direct account,
transforming the RHS. Since $ {\cal H} _ {free} = (m/2) \sum_iv_i^2, v _ {j\alpha} $
commutes  with $ {\cal H} _ {free} $, it is possible to write down
$$ v _ {j\alpha} =e ^ {i {\cal H} _ {free} t ^\prime/\hbar} v _ {j\alpha}
e ^ {-i {\cal H}_{free}t^\prime/\hbar}=v_{j\alpha}(t^\prime),$$ then a bordering $
\exp (i {\cal H} _ {free} t ^\prime/\hbar) \dots -\exp (-i {\cal H} _ {free} t
^\prime/\hbar) $ from the RHS (B24) leaves.  We shall designate the RHS (B24) as $
\Psi_\alpha (t) $, and after calculation of the commutator it appears equal
$$\Psi_\alpha (t) = - {e\over 2}\int_{-\infty}^0dt^\prime\sum_i\left\langle0\left|
v _ {i\beta} {\partial E_\beta ({\bf r} _i, t+t ^\prime) \over \partial r _
{i\alpha}} \right. \right. $$
$$\left.\left. + {\partial E_\beta ({\bf r} _i, t+t ^\prime) \over \partial r _ {i\alpha}}
v _ {i\beta} \right|0\right\rangle. ~~~~~~~~~~~~~~~~~~~ (B25) $$ We shall notice,
that the equality
$$ {1\over 2} \left\langle0\left|v _ {i\beta} {\partial E_\alpha ({\bf r} _i, t) \over
\partial r _ {i\beta}} + {\partial E_\alpha ({\bf r} _i, t) \over
\partial r_{i\beta}}v_{i\beta}\right|0\right\rangle=$$
$$ = (i/\hbar) \langle0 | [{\cal H}, E_\alpha ({\bf r} _i, t)] |0\rangle=0 $$ is carried out
similar (B23), therefore $ \Psi_\alpha (t) $ it is possible to write down as
$$\Psi_\alpha (t) = - {e\over 2}\int_{-\infty}^0dt^\prime\sum_i\left\langle0\left|
v _ {i\beta} \left ({\partial E_\beta ({\bf r} _i, t+t ^\prime) \over
\partial r _ {i\alpha}} \right.\right.\right. $$
$$-\left. {\partial E_\alpha ({\bf r} _i, t+t ^\prime) \over
\partial r _ {i\beta}} \right) + \left ({\partial E_\beta ({\bf r} _i, t+t ^\prime) \over
\partial r _ {i\alpha}} \right. $$
$$\left.\left.-\left. {\partial E_\alpha ({\bf r} _i, t+t ^\prime) \over
\partial r_{i\beta}}\right)v_{i\beta}\right|0\right\rangle_{free},~~~~~~~~~~~~~(B26)$$
that is equal
$$\Psi_\alpha (t) = - {e\over 2}\int_{-\infty}^0dt^\prime\sum_i\langle0|
( {\bf v} _i\times rot \, {\bf E} ({\bf r} _i, t+t ^\prime)) _ \alpha $$
$$- (rot \, {\bf E} ({\bf r} _i, t+t ^\prime) \times {\bf v} _i) _ \alpha|0\rangle.
~~~~~~~~~~~~~~~~~~~ (B27) $$ Using the Maxwell equation $rot \, {\bf E} ({\bf r}, t)
= - (1/c) (\partial {\bf H} ({\bf r}, t) /\partial t) $ and having calculated
integral on $t^\prime $, we obtain, that $ \Psi_\alpha (t) $ is equal to the LHS
(B24), as it was required to prove.

Let us notice, that in a case $ {\bf E} =0, \; {\bf H} =const $, both sides of (B24)
address in 0. It is obvious to the RHS, and in the LHS  the matrix elements appear $
\langle0|v _ {i\beta} |0\rangle = (i/\hbar) \langle0 | [{\cal H}, r _ {i\beta}]
|0\rangle=0 $ because the operator $ {\cal H} $ is diagonal.


\begin{references}
\bibitem {1} H. Stolz. Time Resolved Light Scattering from Exitons. Springer Tracts
in Modern Physics. Springer, Berlin (1994).

\bibitem {2} J. Shah. Ultrafast Spectroscopy of Semiconductors and Semiconductor
Nanostructures. Springer-Verlag, Berlin (1996).

\bibitem {3} R. Zeyher, H. Bilz, M. Cardona. Solid State Commun.
v. 19, pp 57-60, 1967.

\bibitem {4}  L. I. Korovin, I. G. Lang, D. A. Contreras-Solorio, S. T. Pavlov.
Physics of the Solid State (St. Petersburg),  {\bf 43}, N 11, pp.
2182-2191, 2001,; cond-mat/0104262.

\bibitem {5} L.D. Landau, E. M. Lifshitz.
 The field theory, Science, Moscow, 1973, page.
68.
\bibitem {6} L.D. Landau, E. M. Lifshitz. The quantum mechanics, Science, Moscow, 1974, page.
520.

\bibitem {7} A. A. Abrikosov, L. P. Gor'kov, A. E. Dzyaloshinsky.
Methods of the quantum field theory in statistical physics, Moscow, 1998.

\bibitem {8} R. Kubo. J. Phys. Soc. Japan, v. 12, N 6, 570, 1957.

\bibitem {9} S. Nakajima. Proc. Phys. Soc., v. 69, 441, 1956.

\bibitem {10} O. V. Konstantinov, V. I. Perel.  Zh. Eksp. i Teor. Fiz.,
 v. 37, N 3 (9), 786-792, 1959.

\bibitem {11} R. Enderlein, K. Peuker, F. Bechstedt. Phys. stat. solidi (b),
v. 92, 149-158, 1979.

\bibitem {12} D. A. Contreras-Solorio, S. T. Pavlov, L. I. Korovin, I. G. Lang.
Phys. Rev. {\bf B62}, 23, 16815 (2000); cond-mat/0002229.

\bibitem {13}  L. I. Korovin, I. G. Lang, D. A. Contreras-Solorio, S. T. Pavlov
Physics of the Solid State (St. Petersburg), {\bf 44}, N 9, 1681-
1689, 2002; cond-mat/0203390.

\bibitem {14}  L. I. Korovin, I. G. Lang, D. A. Contreras-Solorio, S. T. Pavlov
Physics of the Solid State (St. Petersburg), {\bf 44}, N 11, 2084
-2096, 2002; cond-mat/0104262.
\end{references}
\end{document}